\documentclass[12pt]{iopart}

\usepackage{iopams}  

\usepackage{amsfonts}
\usepackage{amssymb}
\usepackage{bbm}
\usepackage{cancel} 
\usepackage{youngtab}
\usepackage{graphicx}
\usepackage{multicol}

\expandafter\let\csname equation*\endcsname\relax
\expandafter\let\csname endequation*\endcsname\relax

\usepackage{amsmath}
\usepackage{etoolbox}

\patchcmd{\subequations}{\alph{equation}}{\textit{\alph{equation}}}{}{}

\usepackage{cleveref}

\usepackage[caption=false]{subfig}
\usepackage{float}
\usepackage{color}

\begin{document}

\title[From Peccei Quinn symmetry to mass hierarchy problem]{From Peccei Quinn symmetry to mass hierarchy problem}

\author{Y.A. Garnica$^1$, S. F. Mantilla$^2$ and 
R. Martinez$^3$}

\address{$^1$ Departamento de F\'isica, DCI, Campus Le\'on, Universidad de Guanajuato, 37150, Le\'on, Guanajuato, M\'exico}
\address{$^2$ Max-Planck Institute for the Physics of Complex Systems, N\"{o}thnitzer Stra\ss e 38, 01187 Dresden, Germany}
\address{$^3$ Universidad Nacional de Colombia, Departamento de F\'isica, K. 45 No. 26-85, Bogot\'a, Colombia}
\ead{\mailto{ya.garnicagarzon@ugto.mx}, \mailto{mantilla@pks.mpg.de}, \mailto{remartinez@unal.edu.co}}
\vspace{10pt}

\begin{abstract}
We propose a non-universal $\mathrm{U}(1)_{X}$ gauge extension to the Standard Model (SM) and an additional  Peccei-Quinn (PQ) global symmetry to study the mass hierarchy and strong CP problem. The scheme allows us to distinguish among fermion families and to generate the fermionic mass spectrum of particles of the SM. The symmetry breaking is performed by two scalar Higgs doublets and two scalar Higgs singlets, where one of these has the axion which turns out to be a candidate for Cold Dark Matter. The exotic sector is composed by one up-like $T$ and two down-like $J^{1,2}$ heavy quarks, two heavy charged leptons $E,\mathcal{E}$, one additional right-handed neutrino per family $\nu_{R}^{e,\mu,\tau}$, and an invisible axion $a$. In addition, the large energy scale associated to the breaking of the PQ-symmetry gives masses to the right-handed neutrinos in such a way that the active neutrinos acquire eV-mass values due to the see-saw mechanism. On the other hand, from the non-linear effective Lagrangian, the flavour changing of the down quarks and charged leptons with the axion are considered.
\end{abstract}

%
%
%
%
%

\section{Introduction}
The detection of a CP-even scalar field by ATLAS \cite{Atlas_Higgs} and CMS \cite{CMS_Higgs} detectors with a mass of $125 \mathrm{GeV}$ completed the electroweak spontaneous symmetry breakdown (SSB) and mass generation in the Standard Model (SM) \cite{SM}.
The detection of the Higgs boson suggests the possibility to consider additional scalar fields. Some extensions broadly studied are the well-known Two-Higgs-Doublet Models (2HDM) and those ones considering the addition of a singlet scalar field \cite{2HDM}. The 2HDM was proposed to understand the mass differences between the top and bottom quark \cite{2HDM_phen} by considering two vacuum expectation values (VEV). One extension considers the existence of an additional scalar singlet, in this case we have the Next-to-Minimal model (N2HDM) \cite{N2HDM}, where a new scale is associated to its VEV. This is useful when an additional $\mathrm{U}(1)$ local symmetry is considered \cite{MartinezU1}, in which the imaginary part corresponds to the would-be Goldstone boson of the new gauge boson. These models with additional $\mathrm{U}(1)$ gauge symmetry have been extensively studied to understand flavour physics \cite{Flavour}, neutrino physics \cite{Neutrinos}, dark matter (DM) \cite{DarkMatter}, among other phenomena \cite{Other}. The new neutral gauge boson of $\mathrm{U}(1)$ has many phenomenological consequences \cite{Neutralino}. However, in order to get the new $\mathrm{U}(1)$ free of chiral anomalies, it is required to extend the fermionic spectrum accordingly as well as an extended scalar sector to generate the symmetry breakdown to yield heavy masses for the new gauge boson and the exotic fermions. Non universal $\mathrm{U}(1)_{X}$  models in the quark sector have been proposed in \cite{MartinezU1}, giving texture matrices with zeros that suggest a hierarchy among their eigenvalues. Phenomenological consequences of these models have been studied in \cite{Martinez2014}, \cite{Martinez2015}, \cite{Martinez2016}, including the scalar sector to study effects of DM \cite{LangackerU12009}.

From the phenomenological point of view, it is possible to describe some features of the mass hierarchy by assuming Yukawa matrices with zero textures \cite{Hierarchy0}. For instance, the addition of discrete symmetries to the SM are able to yield the Fritzsch Ansatz \cite{Fritzsch1978} which can describe the mass spectrum in the quark sector as well as the CP violation phase observed in the experiments. These mass structures can also be obtained in the lepton sector, as Fukugita, Tanimoto and Yanagida showed in \cite{Fukugita1993}, in which the small masses of neutrinos are understood with the seesaw mechanism with Majorana right-handed neutrinos. Additionally, these neutrinos are able to induce matter-antimatter asymmetry through leptogenesis \cite{Leptogenesis}. Lastly, the mixing in the neutral leptonic sector predicts neutrino flavor oscillations, whose main parameters are the square mass differences of the left-handed light neutrinos and the mixing angles. The confirmation of neutrino oscillations are confirmed by detectors of solar neutrinos such as Homestake \cite{Homestake}, SAGE \cite{SAGE}, GALLEX \& GNO \cite{GALLEX_GNO}, SNO \cite{SNO}, Borexino \cite{Borexino} and Super-Kamiokande \cite{SKamiokande}; atmospheric neutrinos as in IceCube \cite{Icecube}; neutrinos from reactors as KamLAND \cite{KamLAND}, CHOOZ \cite{CHOOZ}, Palo Verde \cite{PaloVerde}, Daya Bay \cite{DayaBay}, RENO \cite{RENO}, and
SBL \cite{SBL}; and neutrinos from accelerators as in MINOS \cite{MINOS}, T2K \cite{T2K}, and NO$\nu$A \cite{NOVA}.

The small masses of the neutrinos, without using fine tuning, can be obtained by two methods: using radiative corrections \cite{RadiativeNeutrino} or through the seesaw mechanism \cite{seesaw}. The seesaw mechanism is implemented by introducing a high energy scale that violates the leptonic number and gives masses to the right-handed neutrinos, and the mixture at the electroweak scale of these with the left-handed neutrinos. The experimental data are compatible with the hypothesis that at least two species of neutrinos have mass, where the left-handed neutrinos are linear combinations of mass eigenstates, and the mixing angles are given by the Pontecorvo-Maki-Nakagawa-Sakata matrix (PMNS) \cite{PMNS1}, \cite{PMNS2}. The parameters are available in NuFIT \cite{NuFIT}.

There are also other problems of fine tuning such as strong CP violation and naturalness problem of the Higgs field mass. To address these problems it is necessary to introduce new physics beyond the SM. Supersymmetric models allow to explain the cancellation of quadratic divergences of the Higgs field mass \cite{Masina:2013wja}. On the other hand, the strong CP problem associated to the anomalous $\mathrm{U}(1)_{A}$ axial symmetry of the QCD \cite{U1Problem} states that the $\theta$-term \cite{Weinberg1978, Wilczek1978} is bounded by the neutron electrical dipole moment \cite{NEDM} and must be less than $10^{-10}$ \cite{Theta_limit}. 
Through a global PQ symmetry \cite{PQ} the fine tuning of strong CP problem can be solved. 
However, the spontaneous global symmetry breaking produces a Nambu-Goldstone boson known as axion that must be invisible because, if the axion would be coupled to the SM fermions, it should had been produced in the colliders. 
Additionally, the coupling of the axion to FCNC is highly suppressed by the processes such as $K\to\pi a$ \cite{flavourdecay,NA62,NA63,ORKA,KOTO}  and $\mu\to ea$ \cite{muea}. 
One way to make the axion invisible is to assume that there is an additional exotic quark with non-vanishing PQ charge so that the axion is disconnected from the low energies physics \cite{Invisibleaxion}. 
Different variations of this model considering 2HDM allow to build  other models with invisible axion \cite{2HDM_axion}. 

From the cosmological considerations \cite{AxionCosmology}, the scale of symmetry breaking must be of the order of $10^{7}-10^{11} \mathrm{GeV}$. 
An interesting framework is that the breaking scale of the PQ symmetry coincides with the scale that gives masses to the right-handed neutrinos necessary to generate the active neutrino masses through seesaw mechanism. 
In this way, the solution of the strong CP problem is then related to the generation of masses for light neutrinos \cite{Neutrino_Axion}. This scale also coincides with the axion decay constant value to have an axion Cold Dark Matter candidate (CDM) \cite{kawasaki}.

If the PQ symmetry is broken after the inflation the topological defects are formed as strings and domain walls which can affect the Universe evolution. This happens practically at the same time of  the QCD transition. The decays of these topological defects produce axions which are different of the misalignment mechanism prediction \cite{aligment}. For this case there are three mechanisms to produce axions: misalignment mechanism, global strings and domain wall decays. The axion relic density can be estimated as the sum of these three mechanisms. 
When the PQ symmetry is broken by the VEV of the additional scalar field at high energies (where the axion can be identified with the phase), the 
axion acquires periodic properties due to non perturbative QCD effects \cite{nonperturbative}, the vacuum misalignment causes the coherent oscillation of the axion field and the ratio between energy of the oscillating axions and the critical density of the present Universe is \cite{kawasaki}
\begin{equation}
\Omega_{a, mis}h^2\approx 4.63\times 10^{-3} \left(\frac{f_a}{10^{10} \mathrm{GeV}}\right)^{1.18} 
\end{equation}
where $h$ is a dimensionless parameter related to the Hubble constant as $h=H_{0}/100\frac{km}{s.Mpc}$ and $\Lambda_{QCD}=400\mathrm{MeV}$ was considered. $f_a$ is the axion decay constant which is the order of $v_S$. 
The spontaneous symmetry breaking of the global PQ symmetry induces a potential that produces global strings which generate axions that contribute to the CDM abundance \cite{CDM}. The ratio between present energy of the axions emitted by the strings and the critical present density can be written as \cite{kawasaki}  
\begin{equation}
\Omega_{a, st}h^2\approx (7.3\pm 3.9)\times 10^{-3} {\cal N}_{DW}^2\left(\frac{f_a}{10^{10} \mathrm{GeV}}\right)^{1.18} .
\end{equation}
When the Universe gets temperature the order of $0.1-1 \mathrm{GeV}$ the axions get mass due to QCD effects and the domain walls appear \cite{paredes}. For the KSVZ type models the domain wall number can be calculated as a function of $\mathrm{U}(1)_{PQ}$ anomaly \cite{anomalia}, which yields a value of ${\cal N}_{DW}=4$ for the model introduced in the current article. The ratio of the critical density with energy density of the produce axions is \cite{kawasaki}
 \begin{equation}
\Omega_{a, dec}h^2\approx (5.4\pm 2.1)\times 10^{-3} \left(\frac{f_a}{10^{10} \mathrm{GeV}}\right)^{1.18} .
\end{equation}
The sum of these three contributions gives the total axion abundance and it is lower than observed CDM abundance, i.e., 
$\Omega_{a, tot}h^2 \leq \Omega_{CDM}h^2$ where $\Omega_{CDM}h^2\approx 0.12$ \cite{densidadCDM}. From these conditions is found that \cite{kawasaki}
\begin{equation}
f_a  \lesssim  1.34\times 10^{10} - 9.5\times 10^{9} \mathrm{GeV}.
\end{equation}
The axion mixes with the $\pi^0$ and $\eta$ mesons, since it has their same quantum numbers and takes a mass
given by \cite{masaaxion}
\begin{align}
m_a &=\frac{\sqrt{m_u m_d}}{m_u + m_d} \frac{m_\pi f_\pi}{f_a},\nonumber \\
m_a &  \gtrsim  (8.7 - 6.1) \times 10^{-4} eV.
\end{align}

The fermion mass hierarchy and their mixing angles using a $\mathrm{U}(1)_{X}$ anomaly free gauge symmetry was studied in the reference \cite{Mantilla}. The model contains additionally one exotic up-like quark, two bottom-like quarks, two charged leptons and one right-handed neutrino per family. Three right-handed sterile neutrinos were introduced to implement the inverse seesaw mechanism to understand the squared mass differences of the active neutrinos and the mixing angles. A $Z_{2}$ symmetry was implemented to generate the zeros of the mass matrices of the fermions. In addition, the model contains two Higgs doublets which give masses at tree level to the third generation of fermions, and a scalar singlet to break the $\mathrm{U}(1)_{X}$ gauge symmetry and gives masses to the exotic particles. In the present work, we build a $\mathrm{U}(1)_{PQ}$ global symmetry to replace the ${Z}_2$ symmetry, which generates the necessary zeros in the mass matrices to explain the mass hierarchy and to solve the strong CP-problem. The PQ symmetry breaking scale generates the masses of the right neutrinos, which gives masses to the active neutrinos of order eV by seesaw mechanism. 

\section{$\mathrm{U}(1)_X$ model construction}

The naturalness problem associated to the mass hierarchy in the SM requires the use of new physics. In this context, the addition of a new $\mathrm{U}(1)$ gauge group, with gauge boson $Z'_{\mu}$ and coupling constant $g_{X}$, is appropriate to understand such a hierarchy and then avoid introducing unpleasant fine tuning. This additional $\mathrm{U}(1)_{X}$ is non-universal, i.e., the charge set for each generation is different to the other two. The conditions to generate the cancellation of anomalies under this new group requires cancellation of the following constrains
\begin{align}
 & & \sum_{Q}X_{Q_{L}} - \sum_{Q}X_{Q_{R}}=0,	\nonumber\\
 & & \sum_{\ell}X_{\ell_{L}} + 3\sum_{Q}X_{Q_{L}}=0,\nonumber	\\
& &
	\sum_{\ell, Q}\left[Y_{\ell_{L}}^{2}X_{\ell_{L}}+3Y_{Q_{L}}^{2}X_{Q_{L}} \right] - \sum_{\ell,Q}\left[Y_{\ell_{R}}^{2}X_{L_{R}}+3Y_{Q_{R}}^{2}X_{Q_{R}} \right]=0,\nonumber	\\
 & &
	\sum_{\ell, Q}\left[Y_{\ell_{L}}X_{\ell_{L}}^{2}+3Y_{Q_{L}}X_{Q_{L}}^{2} \right] - \sum_{\ell, Q}\left[Y_{\ell_{R}}X_{\ell_{R}}^{2}+3Y_{Q_{R}}X_{Q_{R}}^{2} \right]=0,\nonumber	\\
& &
	\sum_{\ell, Q}\left[X_{\ell_{L}}^{3}+3X_{Q_{L}}^{3} \right] - \sum_{\ell, Q}\left[X_{\ell_{R}}^{3}+3X_{Q_{R}}^{3} \right]=0,\nonumber		\\
& &
	\sum_{\ell, Q}\left[X_{\ell_{L}}+3X_{Q_{L}} \right] - \sum_{\ell, Q}\left[X_{\ell_{R}}+3X_{Q_{R}} \right]=0,
\end{align}
corresponding to the triangular anomalies 
$\left[\mathrm{\mathrm{SU}(3)}_{C} \right]^{2} \mathrm{\mathrm{U}(1)}_{X}$, $\left[\mathrm{\mathrm{SU}(2)}_{L} \right]^{2} \mathrm{\mathrm{U}(1)}_{X}$,  $\left[\mathrm{\mathrm{U}(1)}_{Y} \right]^{2}   \mathrm{\mathrm{U}(1)}_{X}$, $\mathrm{\mathrm{U}(1)}_{Y}   \left[\mathrm{\mathrm{U}(1)}_{X} \right]^{2}$, $\left[\mathrm{\mathrm{U}(1)}_{X} \right]^{3}$ and $\left[\mathrm{Grav} \right]^{2}\mathrm{\mathrm{U}(1)}_{X}$, respectively.

\begin{table}
\centering
\begin{tabular}{ccc}
Scalar bosons	&	$X$	&	$\mathrm{U}(1)_{PQ}$	\\ \hline 
\multicolumn{2}{c}{Higgs Doublets}\\ \hline\hline
$\phi_{1}=\left(\begin{array}{c}
\phi_{1}^{+} \\ \frac{h_{1}+v_{1}+i\eta_{1}}{\sqrt{2}}
\end{array}\right)$	&	$2/3$	&	$x_1$	\\
$\phi_{2}=\left(\begin{array}{c}
 \phi_{2}^{+} \\ \frac{h_{2}+v_{2}+i\eta_{2}}{\sqrt{2}}
\end{array}\right)$	&	$1/3$	&	$x_2$	\\   \hline\hline
\multicolumn{2}{c}{Higgs Singlets}\\ \hline\hline
$\chi  =\frac{\xi_{\chi}  +v_{\chi}  +i\zeta_{\chi}}{\sqrt{2}}$	& $-1/3$	&	$x_{\chi}$	\\   
$\sigma =\frac{\xi_{\sigma}+v_{\sigma}+i\zeta_{\sigma}}{\sqrt{2}}$	& $-1/3$	&	$x_\sigma$	\\  
$S =\frac{\xi_{S}+v_{S}+i\zeta_{S}}{\sqrt{2}}$	& $0$	&	$x_S$	\\
\hline \hline 
\end{tabular}
\caption{Non-universal $X$ and $\mathrm{U}(1)_{PQ}$ charges for scalar sector. The explicit form of the PQ charges are given by \eqref{eq:AxionHiggs}.}
\label{tab:Scalar-content}
\end{table}

\begin{tiny}
\begin{table}
\centering
\begin{tabular}{ccc|ccc|}
\hline\hline
Quarks	&	$X$	&&	Leptons	&	$X$	\\ \hline 
\multicolumn{5}{c}{SM Fermionic Isospin Doublets}	\\ \hline\hline
$q^{1}_{L}=\left(\begin{array}{c}U^{1} \\ D^{1} \end{array}\right)_{L}$
	&	$+1/3$		&&
$\ell^{e}_{L}=\left(\begin{array}{c}\nu^{e} \\ e^{e} \end{array}\right)_{L}$
	&	$0$		\\
$q^{2}_{L}=\left(\begin{array}{c}U^{2} \\ D^{2} \end{array}\right)_{L}$
	&	$0$			&&
$\ell^{\mu}_{L}=\left(\begin{array}{c}\nu^{\mu} \\ e^{\mu} \end{array}\right)_{L}$
	&	$0$			\\
$q^{3}_{L}=\left(\begin{array}{c}U^{3} \\ D^{3} \end{array}\right)_{L}$
	&	$0$			&&
$\ell^{\tau}_{L}=\left(\begin{array}{c}\nu^{\tau} \\ e^{\tau} \end{array}\right)_{L}$
	&	$-1$		\\   \hline\hline

\multicolumn{5}{c}{SM Fermionic Isospin Singlets}	\\ \hline\hline
\begin{tabular}{c}$U_{R}^{1,2,3}$\\$D_{R}^{1,2,3}$\end{tabular}	&	 
\begin{tabular}{c}$+2/3$\\$-1/3$\end{tabular}		&&
\begin{tabular}{c}$e_{R}^{e,\tau}$\\$e_{R}^{\mu}$\end{tabular}	&	
\begin{tabular}{c}$-4/3$\\$-1/3$\end{tabular}	\\   \hline \hline 

\multicolumn{2}{c}{Non-SM Quarks}	&&	\multicolumn{2}{c}{Non-SM Leptons}	\\ \hline \hline
\begin{tabular}{c}$T_{L}$\\$T_{R}$\end{tabular}	&
\begin{tabular}{c}$+1/3$\\$+2/3$\end{tabular}		&&
\begin{tabular}{c}$\nu_{R}^{e,\mu,\tau}$\\$E_L$\\$E_R$\end{tabular} 	&	
\begin{tabular}{c}$1/3$\\$-1$\\$-2/3$\end{tabular}	\\
$J^{1,2}_{L}$	&	  $0$ 		&&	$\mathcal{E}_{L}$	&	$-2/3$		\\
$J^{1,2}_{R}$	&	 $-1/3$		&&	$\mathcal{E}_{R}$	&	$-1$		\\ \hline \hline
\end{tabular}
\caption{Non-universal $X$ charges for fermionic sector. The PQ charges for the fermions are shown in the Table \ref{tab:PQ-charges} and in the \ref{app:PQ-charges}.}

\label{tab:Fermionic-content}
\end{table}
\end{tiny}

A new $\mathrm{U}(1)_{PQ}$ global symmetry, whose charges are shown in Tables \ref{tab:Scalar-content} and \ref{tab:Fermionic-content} for the scalar and fermion sectors, is introduced in replacement of $Z_{2}$ of previous approaches \cite{Mantilla}. These charges are chosen in such a way that the allowed Yukawa parameters are equal to those ones allowed by the previous $Z_{2}$ discrete symmetry  in \cite{Mantilla}, i.e., the PQ symmetry allows to understand the fermion mass hierarchy and the strong CP-problem. Alike, the scale associated to the SSB of the PQ symmetry is suitable for understanding the small mass scale for active neutrinos through the see-saw mechanism. In this way, the table \ref{tab:Scalar-content} shows the scalar fields associated to the electroweak ($\phi_{1,2}$) and $\mathrm{U}(1)_{X}$ ($\chi$) SSBs. Also, a $\sigma$ scalar field is introduced to give masses to the lighter fermions through radiative corrections, and $S$, a scalar singlet is implemented to break the $\mathrm{U}(1)_{PQ}$ symmetry, generating a pseudo-Goldstone boson that turns out to be an invisible axion \cite{Invisibleaxion} whose mass comes from non-perturbative contributions.

\subsection{Gauge boson masses ($B_\mu, W_\mu^3, Z'_\mu$)}


The $\mathrm{SU}(2)_{L}\otimes\mathrm{U}(1)_{Y}\otimes\mathrm{U}(1)_{X}$ gauge symmetry generates an additional term in the covariant derivative
\begin{align}
    D_{\mu}=\partial_{\mu} +igW_{\mu}^{a}T_{a} + ig'\frac{Y}{2}B_{\mu}+ig_{X}XZ'_{\mu},
    \label{eq:Covariant_derivative}
\end{align}
After the SSBs the $W^{\pm}_{\mu}=(W_{\mu}^{1}\mp  W_{\mu}^{2})/\sqrt{2}$ acquire masses equal to $M_{W}=gv/2$ where $v=\sqrt{v_{1}^{2}+v_{2}^{2}}$. The masses of the $(B_{\mu}, W_{\mu}^{3},  Z_{\mu}^{\prime})$ neutral gauge bosons are obtained from the following matrix

\begin{equation}
M^{2}=\frac{1}{4}\left(
    \begin{array}{ccc}
    g^{2} v^{2} & -gg'v^{2} & \frac{2}{3}g' g_{X} v^{2}(1+s^{2}_{\beta})\\
    -gg'v^{2} & g'{}^{2} v^{2} & -\frac{2}{3}g g_{X} v^{2}(1+s_{\beta}^{2})\\
    \frac{2}{3}g' g_{X} v^{2}(1+s^{2}_{\beta}) & -\frac{2}{3}g g_{X} v^{2}(1+s_{\beta}^{2}) & \frac{4}{9}g_{X}^{2} v_{\chi}^{2}\left[1+(1+3s^{2}_{\beta})\frac{v^{2}}{v_{\chi}^{2}}\right]\\
\end{array} 
\right) ,
\end{equation}

which has one eigenvalue equal to zero corresponding to the photon and two eigenvalues corresponding to the masses of $Z_\mu$, $Z'_\mu$, respectively, given by
\begin{align}
    M_{Z}&\approx\frac{gv}{2\cos{\theta_{W}}},   &M_{Z'}&\approx \frac{g_{X}v_{\chi}}{3}.
    \label{eq:Gauge_Masses}
\end{align}
 The matrix that diagonalized $M$ is given in \cite{Mantilla} and it has the form
\begin{equation}
    R= \left(
    \begin{array}{ccc}
    1&0&0\\
    0&c_{Z}&s_{Z}\\
    0&-s_{Z}&c_{Z}
    \end{array}
    \right)
    \left(\begin{array}{ccc}
    c_{W}&s_{W}&0\\
    -s_{W}&c_{W}&0\\
    0&0&1
    \end{array}\right) ,
\end{equation}
where $t_{W}=g'/g$ is the Weinberg angle and $s_{Z}$ is the sine of the mixing angle between $Z_\mu$ and $Z'_\mu$ gauge bosons
\begin{equation}
    s_{Z}\approx\left(1+s_{\beta}^{2}\right)\frac{2g_{X}c_{W}}{3g}\left(\frac{m_{Z}}{m_{Z'}}\right)^{2}.
\end{equation}

The lower bound for the $Z'$ mass coming from LHC by using sequential SM couplings is of the order of $M_{Z'} \geq 4.1$ TeV \cite{mzLHC} which can be translated as
$v_{\chi} \geq 15/{g_X}$ TeV. Using the couplings of this model 
a bound for $Z'$ mass was gotten as  $M_{Z'}>6\mathrm{\,TeV}$   \cite{ourmodel}, which implies $v_\chi \geq 18/g_X\mathrm{\,TeV}$. In the perturbative limit ($g_X=1$) this bound can be written as  $v_\chi \geq 18\mathrm{\,TeV}$.

In order to define the mass eigenstates associated to the would-be Goldstone bosons of the  $Z_\mu$ and $Z'_\mu$ gauge fields (\ref{eq:Gauge_Masses}), it is necessary to use the bilinear terms $Z_{\mu}\partial^{\mu}G_{Z}$ that come from the kinetic terms of the scalar fields. It is expected that these contributions are canceled out by the bilinear terms originated in the gauge fixing. The gauge fixing Lagrangian has the form
\begin{equation}
    \mathcal{L}_{GF}=-\frac{1}{2}\left(\partial_{\mu}Z^{\mu}+M_{Z}G_{Z}\right)^{2}-\frac{1}{2}\left(\partial_{\mu}Z^{'\mu}+M_{Z'}G_{Z'}\right)^{2}.
\end{equation}
After integrating by parts, the terms in which we are interested in are
\begin{equation}
    M_{Z}Z^{\mu}\partial_{\mu}G_{Z}+ M_{Z'}Z'^{\mu}\partial_{\mu}G_{Z'},
\end{equation}
which are expected to be canceled out by the contributions coming from covariant derivatives. In order to get the Goldstone boson mass eigenstates, it is necessary to rotate the covariant derivative in function of the mass eigenstates as follows
\begin{align}
   D_{\mu}=\partial_{\mu}&-\left(\frac{ig}{c_{W}}c_{Z}\left(T_{3L}-s_{W}^{2}Q\right)+\frac{g_{X}}{g}c_{W}s_{Z}X\right)Z_{\mu}\nonumber\\ 
   &-ig_{X}\left(-\frac{g}{g_{X}}\frac{s_{Z}}{c_{W
   }}\left(T_{3L}-s_{W}^{2}Q\right)+c_{Z}X\right)Z'_{\mu}.
\end{align}
The covariant derivative is then acted on $\phi_{1}$, $\phi_{2}$ and $\chi$ to get the following terms associated to the neutral components
\begin{align}
(D_{\mu}\phi_{1})(D^{\mu}\phi_{1})^{\dagger}\approx&-v_{1}Z_{\mu}\partial_{\mu}\eta_{1}\left(-\frac{g}{2c_{W}}c_{Z}+\frac{2g_{X}}{3}s_{Z}\right)\nonumber\\
&-v_{1}Z'_{\mu}\partial_{\mu}\eta_{1}\left(\frac{g}{2c_{W}}s_{Z}+\frac{2g_{X}}{3}c_{Z}\right),\\
 (D_{\mu}\phi_{2})(D^{\mu}\phi_{2})^{\dagger}\approx&-v_{2}Z_{\mu}\partial_{\mu}\eta_{2}\left(-\frac{g}{2c_{W}}c_{Z}+\frac{g_{X}}{3}s_{Z}\right)\nonumber\\
 &-v_{2}Z'_{\mu}\partial_{\mu}\eta_{2}\left(\frac{g}{2c_{W}}s_{Z}+\frac{g_{X}}{3}c_{Z}\right),\\
 (D_{\mu}\chi)(D^{\mu}\chi)^{\dagger}\approx&-v_{\chi}Z_{\mu}\partial_{\mu}\zeta_{X}g_{X}\left(-\frac{1}{3}\right)s_{Z}\nonumber\\
 &-v_{\chi}Z'_{\mu}\partial_{\mu}\zeta_{X}g_{X}\left(-\frac{1}{3}\right)c_{Z}.
\end{align}

Thus, by matching the contributions of the covariant derivatives with the bilinear terms from the gauge fixing Lagrangian, and by replacing $M_{Z}$ and $M_{Z'_{\mu}}$, we obtain the would-be Goldstone bosons associated to the massive neutral gauge bosons
\begin{align}
G_{Z}&=\frac{2c_{W}c_{Z}}{gv}\left[\frac{g}{2c_{W}}\left(v_{1}\eta_{1}+v_{2}\eta_{2}\right)\right]+\frac{2c_{W}s_{Z}}{gv}\left[-\frac{g_{X}}{3}\left(2v_{1}\eta_{1}+v_{2}\eta_{2}\right)\right]\nonumber\\
 &+\frac{2c_{W}}{gv}v_{\chi}\zeta_{\chi}\frac{g_{X}}{3}s_{Z},\\
G_{Z'}&=\frac{3}{g_{\chi}v_{\chi}}\left[\frac{g}{2c_{W}}s_{Z}\left(-v_{1}\eta_{1}-v_{2}\eta_{2}\right)+\frac{g_{\chi}c_{Z}}{3}\left(-2\frac{v_{1}}{v_{\chi}}\eta_{1}-\frac{v_{2}}{v_{\chi}}\eta_{2}\right)\right]\nonumber\\
 &
 +\frac{3}{g_{\chi}v_{\chi}}\left(v_{\chi}\zeta{\chi}\frac{g_{X}}{3}c_{Z}\right).
\end{align}
Under the approximation $s_{Z}\sim 0,c_{Z}\sim 1$, it is possible to write
\begin{align}
    G_{Z}&\approx s_{\beta}\eta_{1}+c_{\beta}\eta_{2}+\frac{M_{Z'}}{M_{Z}}s_{Z}\zeta_{\chi},\nonumber\\
    G_{Z'}&\approx\zeta_{\chi}-2\frac{v_{1}}{v_{\chi}}\eta_{1}-\frac{v_{2}}{v_{\chi}}\eta_{2}.
\label{eq:eaten-gauge-phases}
\end{align}
The definition of would-be Goldstone bosons allow us to impose new conditions for PQ charges in order to decouple the axion.

\subsection{Yukawa Lagrangian densities}
Regarding the Yukawa sector, the two Higgs doublets $\phi_{1,2}$ are required to generate the masses of the top quark $t$ through $v_{1}$, and $\phi_{2}$ does the same through $v_{2}$ to the bottom quark $b$ as well as the $\mu$ and $\tau$ leptons at tree level. On the other hand, the $\chi$ singlet provides mass to the exotic charged particles at $\mathrm{U}(1)_{X}$ breaking scale, i.e., the order of the $Z'$ mass, which has to be larger than the TeV scale according  to the LHC restrictions \cite{ATLAS}. 

The Yukawa Lagrangian densities used in this article are taken from \cite{Mantilla} in order to get the appropriate zero textures, formerly obtained with a $Z_{2}$ discrete symmetry. In the present article, instead of the $Z_{2}$, the $\mathrm{U(1)}_{PQ}$ plays the role of producing the zero textures, so that, the set of PQ charges are obtained such that the same Yukawa Lagrangians are invariant under the action of $\mathrm{U(1)}_{PQ}$. Then, the Yukawa Lagrangian for the quark sector is given by
\begin{align}
 \label{eq:Lagrangian-quark-sector}
 -\mathcal{L}_Q &=
\overline{q_L^{1}}\left(\widetilde{\phi}_{2}h^{U}_{2}\right)_{12}U_{R}^{2}
+\overline{q_L^{1}}\left(\widetilde{\phi}_{2}h^{T}_{2} \right)_{1}T_{R}
+\overline{q_L^{2}}(\widetilde{\phi}_{1} h^{U}_{1})_{22}U_R^{2}
+\overline{q_{L}^{2}} (\widetilde{\phi}_{1} h^{T}_{1})_{2}T_{R}
\nonumber\\
&
+\overline{q_L^{3}}(\widetilde{\phi}_{1} h^{U}_{1})_{31}U_{R}^{1}
+\overline{q_L^{3}}(\widetilde{\phi}_{1}h^{U}_{1})_{33}U_R^{3}
+\overline{T_{L}}\left(\chi h_{\chi}^{U}\right)_{2}{U}_{R}^{2}
+\overline{T_{L}}\left(\sigma h_{\sigma}^U\right)_{1,3}U_{R}^{1,3}\\
&
+\overline{T_{L}}\left(\chi h_{\chi}^{T}\right)T_{R}
+\overline{q_L^{1}} (\phi_{1}h^{J}_{1})_{1n} J^{n}_{R}
+\overline{q_L^{2}}\left(\phi_{2}h^{J}_{2}\right)_{2n} J^{n}_{R}
+\overline{q_L^{3}}\left(\phi_{2}h^{D}_{2} \right)_{3j}D_R^{j} 
\nonumber\\
&
+\overline{J_{L}^n}\left(\sigma^{*}h_{\sigma }^{D}\right)_{nj}{D}_{R}^{j}
+\overline{J_{L}^{n}}\left(\chi ^*h_{\chi }^{J}\right)_{nn}{J}_{R}^{n}+h.c., \nonumber
\end{align}
(with $n=1,2$ and $j=1,2,3$). On the other hand, the Yukawa Lagrangian for the lepton sector is given by

\begin{align}
\label{eq:Leptonic-Lagrangian}
-\mathcal{L}_{L} &= 
g_{e\mu}^{2e}\overline{\ell^{e}_{L}}\phi_{2}e^{\mu}_{R} 
+g_{\mu\mu}^{2e}\overline{\ell^{\mu}_{L}}\phi_{2}e^{\mu}_{R} 
+g_{\tau e}^{2e}\overline{\ell^{\tau}_{L}}\phi_{2}e^{e}_{R}
+g_{\tau\tau}^{2e}\overline{\ell^{\tau}_{L}}\phi_{2}e^{\tau}_{R} 
\nonumber\\
&
+g_{Ee}^{1}\overline{\ell^{e}_{L}}\phi_{1}E_{R} 
+g_{E\mu}^{1}\overline{\ell^{\mu}_{L}}\phi_{1}{E}_{R}
+h^{\sigma e}_{E}\overline{E_{L}}\sigma^{*} e^{e}_{R} 
+ h^{\sigma\mu}_{\mathcal{E}}\overline{\mathcal{E}_{L}}\sigma e^{\mu}_{R} \nonumber\\
&
+ h^{\sigma\tau}_{E}\overline{E_{L}}\sigma^{*} e^{\tau}_{R}
+ h^{\chi E}\overline{E_{L}}\chi E_{R}
+h^{\chi\mathcal{E}}\overline{\mathcal{E}_{L}}\chi^{*} \mathcal{E}_{R}
+h_{2e}^{\nu i}\overline{\ell^{e}_{L}}\tilde{\phi}_{2}\nu^{i}_{R} \nonumber\\
&+ h_{2\mu}^{\nu i}\overline{\ell^{\mu}_{L}}\tilde{\phi}_{2}\nu^{i}_{R}
+h_{S i}^{\nu j} \overline{\nu_{R}^{i\;C}} S\nu_{R}^{j},
\end{align}
with $i,j=e,\mu,\tau$. In contrast to \cite{Mantilla}, the current model does not introduce Majorana fields $N_{R}^{e,\mu,\tau}$ since they do not mix to the other neutral fermions, making them irrelevant in this approach. Instead, the three $\nu_{R}^{e,\mu,\tau}$ get masses through the VEV of the $S$ scalar field at PQ scale, which is enough to perform the see-saw mechanism. The mass matrices produced by these Lagrangians are studied more profoundly in section \ref{sect:Mass-Matrices}.

The set of the PQ charges are the solutions of a set of coupled algebraic equations obtained from the Yukawa Lagrangians and the scalar potential shown below in \eqref{eq:scalar-potential}. The details are shown in the \ref{app:PQ-charges}. It is important to remark that these equations modify slightly the Yukawa Lagrangian for the quark sector in  \eqref{eq:Lagrangian-quark-sector} as follows. The terms involving $D_{R}^{j}$, i.e., $\overline{q_L^{3}}\left(\phi_{2}h^{D}_{2} \right)_{3j}D_R^{j}$ and $\overline{J_{L}^n}\left(\sigma^{*}h_{\sigma }^{D}\right)_{nj}{D}_{R}^{j}$ were allowed for $j=1,2,3$ since the $D_{R}^{j}$ were universal under $Z_{2}$ in the previous paper \cite{Mantilla}, although the solution of the equations that yield the PQ charge set breaks the universality and allows only the case $j=3$. 
\subsection{Scalar Lagrangian and mass spectrum for the scalar sector}
The most general scalar potential in the PQ symmetry scenario is
\begin{align}
\label{eq:scalar-potential}
V &= \mu_{1}^{2}\phi_{1}^{\dagger}\phi_{1} 
+ \mu_{2}^{2}\phi_{2}^{\dagger}\phi_{2} 
+ \mu_{\chi}^{2}\chi^{*}\chi 
+ \mu_{\sigma}^{2}\sigma^{*}\sigma
+\mu_{S}^2 S^{*}S
+ \lambda_{1}\left(\phi_{1}^{\dagger}\phi_{1}\right)^{2} \nonumber\\
&
+ \lambda_{2}\left(\phi_{2}^{\dagger}\phi_{2}\right)^{2} 
+ \lambda_{3}\left(\chi^{*}\chi \right)^{2} 
+ \lambda_{4}\left(\sigma^{*}\sigma \right)^{2} 
+ \lambda_{5}\left(\phi_{1}^{\dagger}\phi_{1}\right) \left(\phi_{2}^{\dagger}\phi_{2}\right) \nonumber\\
&+ \lambda'_{5}\left(\phi_{1}^{\dagger}\phi_{2}\right)\left(\phi_{2}^{\dagger}\phi_{1}\right)
+ \left(\phi_{1}^{\dagger}\phi_{1}\right)\left[\lambda_{6}\left(\chi^{*}\chi \right)+ \lambda'_{6}\left(\sigma^{*}\sigma \right) \right]
\\
&
+ \left(\phi_{2}^{\dagger}\phi_{2}\right)\left[ \lambda_{7}\left(\chi^{*}\chi \right) + \lambda'_{7}\left(\sigma^{*}\sigma \right) \right] 
+ \lambda_{8}\left(\chi^{*}\chi \right)\left(\sigma^{*}\sigma \right) 
+\lambda_{9}(S^{*}S)^2\nonumber\\
&
+ (S^{*}S)\left[\lambda_{10}\left(\phi_{1}^{\dagger}\phi_{1}\right)+\lambda_{11}\left(\phi_{2}^{\dagger}\phi_{2}\right)+\lambda_{12}\left(\chi^{*}\chi\right)+\lambda_{13}\left(\sigma^{*}\sigma\right)\right]  \nonumber\\                               &+\lambda_{14}\left(\chi^* S\phi_{1}^{\dagger}\phi_{2}+h.c.\right),\nonumber
\end{align}
where the term proportional to $\lambda_{14}$ is needed to avoid trivial PQ charges for the scalar sector. 

\subsubsection{Charged scalar}
For the charged scalar sector, we have the rank 1 matrix in the basis $(\phi_{1}^{\pm},\phi_{2}^{\pm})$
\begin{equation}
    M_{C}^{2}=\frac{1}{4}\left(
    \begin{array}{cc}
    \lambda_{5}^{'}v_{2}^{2}-\lambda_{14}\frac{v_{2}v_{\chi}v_{S}}{v_{1}}&\lambda_{5}^{'}v_{1}v_{2}+\lambda_{14}v_{\chi}v_{S}\\
    *&\lambda_{5}^{'}v_{1}^{2}-\lambda_{14}\frac{v_{1}v_{\chi}v_{S}}{v_{2}}    \end{array}
    \right),
\end{equation}
that implies one would-be Goldstone boson $G_{W}^{\pm}$ associated to the $W_{\mu}^{\pm}$ boson and one charged physical scalar $H^{\pm}$ with mass given by
\begin{equation}
    m_{H^{\pm}}^{2}=\frac{1}{2}\left(\lambda_{5}^{'}v^{2}-\lambda_{14}\frac{2v_{\chi}v_{S}}{s_{2\beta}}\right),
\end{equation}
where $t_{\beta}=\tan\beta=v_{1}/v_{2}$ is the tangent of the mixing angle, $s_{\beta}=\sin\beta$ and $c_{\beta}=\cos\beta$. 


\subsubsection{CP-even scalar}
After the SSB the two Higgs doublets and the singlets acquire VEVs, yielding the mass matrix $M_{R}^{2}$ for CP-even scalar particles expressed in the $(h_{1},h_{2},\xi_{\chi},\xi_{S})$ basis
\begin{equation}
M_{R}^{2}=
    \left(
\begin{array}{cc}
 \lambda_{1}v_{1}^{2}-\frac{\lambda_{14}}{4}\frac{v_{2}v_{\chi}v_{S}}{v_{1}}&\frac{\bar{\lambda}_{5}}{2}v_{1}v_{2}+\frac{\lambda_{14}}{4}v_{\chi}v_{S}\\
 *&\lambda_{2} v_{2}^{2}-\frac{\lambda_{14}}{4}\frac{v_{1}v_{\chi}v_{S}}{v_{2}}\\
 *&*\\
 *&*\\
\end{array} 
\right. \\
\left.
\begin{array}{cccc}
\frac{\lambda_{6}}{2}v_{1}v_{\chi}+\frac{\lambda_{14}}{4}v_{2}v_{S}&\frac{\lambda_{14}}{4}v_{2}v_{\chi}+\frac{\lambda_{10}}{2}v_{1}v_{S}\\
\frac{\lambda_{7}}{2}v_{2}v_{\chi}+\frac{\lambda_{14}}{4}v_{1}v_{S}&\frac{\lambda_{14}}{4}v_{1}v_{\chi}+\frac{\lambda_{11}}{2}v_{2}v_{S}\\
\lambda_{3}v_{\chi}^{2}-\frac{\lambda_{14}}{4}\frac{v_{1}v_{2}v_{S}}{v_{\chi}}&\frac{\lambda_{14}}{4}v_{1}v_{2}+\frac{\lambda_{12}}{2}v_{\chi}v_{S}\\
*&\lambda_{9}v_{S}^{2}-\frac{\lambda_{14}}{4}\frac{v_{1}v_{2}v_{\chi}}{v_{S}}\\
\end{array}
\right) \, ,
\end{equation}
where $\bar{\lambda_{5}}=\lambda_{5}+\lambda'_{5}$. Since $\mathrm{Rank}(M_{R}^{2})=4$, so the four CP-even scalars acquire mass at different scales given by the VEV hierarchy $v_{S}\gg v_{\chi}\gg v \gg v_{\sigma}$. The $v_{\sigma}$, which is required to give radiative masses to the lightest fermions, is small and then it is not considered in the mass matrices of the scalar sector. In addition, in order to decouple the SM from the exotic sectors, the following relations on the scalar coupling constants are assumed to build a natural hierarchy between the PQ and electroweak scale, without unpleasant fine tuning \cite{Bertolini_Santamaria}
\begin{align*}
    \lambda_{6}&\equiv a_{6}\frac{v_{1}^{2}}{v_{\chi}^{2}},\quad
    \lambda_{7}\equiv a_{7}\frac{v_{2}^{2}}{v_{\chi}^{2}},\quad
    \lambda_{10}\equiv a_{10}\frac{v_{1}^{2}}{v_{S}^{2}},\quad
    \lambda_{11}\equiv a_{11}\frac{v_{2}^{2}}{v_{S}^{2}}, \\
    \lambda_{12}&\equiv a_{12}\frac{v_{\chi}^{2}}{v_{S}^{2}},\quad
     \lambda_{14}\equiv a_{14}\frac{v^{2}}{v_{\chi}v_{S}}.
\end{align*}
By implementing such relations, the leading-order contribution to the $M_{R}^{2}$ matrix is given by 
\begin{equation}
\mathit{M}_{\mathrm{R}}^{2} \! \approx \!\!\!
\left(
\begin{array}{cccc}
\lambda_{1}v_{1}^{2}-\frac{a_{14}}{4}\frac{v^{2}v_{2}}{v_{1}}&\frac{\bar{\lambda_{5}}}{2}v_{1}v_{2}+\frac{a_{14}}{4}v^{2}&0&0\\\frac{\bar{\lambda_{5}}}{2}v_{1}v_{2}+\frac{a_{14}}{4}v^{2}&\lambda_{2}v_{2}^{2}-\frac{a_{14}}{4}\frac{v^{2}v_{1}}{v_{2}}&0&0\\
0&0&\lambda_{3}v_{\chi}^{2}&0\\0&0&0&\lambda_{9}v_{S}^{2}
\end{array}
\right),
\label{eq:CP-even-approx}
\end{equation}
yielding that largest eigenvalues are decoupled from the electroweak scale. The remaining $2\times 2$ subblock at electroweak scale is diagonalized by the orthogonal transformation
\begin{align} 
\left(
\begin{array}{c}
h_{SM} \\ H
\end{array}
\right)= 
\left(
\begin{array}{cc}
\cos\theta & \sin\theta \\
-\sin\theta &\cos\theta
\end{array}
\right)
\left(
\begin{array}{c}
h_1 \\ h_2
\end{array}
\right),
\end{align}
where $h_{SM}$ corresponds to the Higgs boson of the SM. The mixing angle between $h_1$ and $h_2$ CP is
\begin{equation}
\tan 2\theta\approx \frac{2(\bar{\lambda}_{5}s_{2\beta}+a_{14})t_{\beta}}{a_{14}(1-t_{\beta}^{2})+2(\lambda_{2}-\lambda_{1}t_{\beta}^{2})s_{2\beta}}.
\end{equation}
It was assumed for the characteristic polynomial of the matrix (\ref{eq:CP-even-approx}) that $\tan\beta\gg 1$, $\sin\beta\approx 1$ and consequently $\cos\beta\approx 0$ since $v_{1}\gg v_{2}$ in order to achieve the top-bottom quark mass hierarchy. 
Then, the corresponding masses for the CP-even sector are
\begin{equation}
m_{h_{SM}}^{2}\approx \lambda_{1}v_{1}^{2},\quad		
m_{H}^{2}\approx-\frac{a_{14}v^{2}t_{\beta}}{4},	\quad
m_{H_{\chi}}^{2}\approx \lambda_{3}v_{\chi}^{2},	\quad
m_{H_{S}}^{2}\approx \lambda_{9}v_{S}^{2},
\end{equation}
where the first and smallest one corresponds to the Higgs boson with mass given by $\lambda_{1}v_{1}^{2}\approx (125\mathrm{GeV})^{2}$, and the conditions $t_{\beta}=v_{1}/v_{2}, \lambda_{1}, \lambda_{3}, \lambda_{9}>0$ and $a_{14}<0$ to fulfill positive eigenvalues. 
\subsubsection{CP-odd scalars}

The CP-odd scalar mass matrix in the $(\eta_{1},\eta_{2},\zeta_{\chi},\zeta_{S})$ basis is given by
\begin{equation}
    M_{I}^{2}=-\frac{\lambda_{14}}{4}
    \left(
    \begin{array}{cccc}
    \frac{v_{2}v_{\chi}v_{S}}{v_{1}}&-v_{\chi}v_{S}&-v_{2}v_{S}&v_{2}v_{\chi}\\
    -v_{\chi}v_{S}&\frac{v_{1}v_{\chi}v_{S}}{v_{2}}&v_{1}v_{S}&-v_{1}v_{\chi}\\
    -v_{2}v_{S}&v_{1}v_{S}&\frac{v_{1}v_{2}v_{S}}{v_{\chi}}&-v_{1}v_{2}\\
    v_{2}v_{\chi}&-v_{1}v_{\chi}&-v_{1}v_{2}&\frac{v_{1}v_{2}v_{\chi}}{v_{S}}
     \end{array}
     \right),
     \label{Cp-odd_sector}
\end{equation}
where Rank($M_{I}^{2})=1$. Thus, there are three zero modes corresponding to $G_{Z}$, $G_{Z'}$ and the axion associated to the SSB of $\mathrm{U}(1)_{PQ}$, which obtain its mass through non-perturbative QCD corrections. The non-vanishing mass eigenvalue is associated to the $A^{0}$ pseudoscalar boson, given by
\begin{equation}
    m_{A^0}^{2}=-\frac{a_{14}v^{2}t_{\beta}}{4},
\end{equation}
where the VEV hierarchy had been implemented.

The would-be Goldstone bosons in (\ref{eq:eaten-gauge-phases}) are useful to get the expressions for the axion $a$ and the physical CP-odd scalar $A^{0}$ through the Gram-Schmidt method. In this way, by considering the hierarchy among VEVs, $v_S\gg v_\chi\gg v_1, v_2$, the orthogonal transformation that goes from the basis $\{\eta_1, \eta_2, \zeta_{\chi},\zeta_{S}\}$ to the mass eigenbasis $\{G_{Z}, G_{Z'}, a, A^0\}$ is given by 
\begin{equation}
R_I=
\left(
    \begin{array}{cccc}
   s_\beta& c_\beta & (1+s^2_\beta)\frac{v}{v_\chi} & 0\\
   -\frac{2v_1}{v_\chi} & -\frac{v_2}{v_\chi} & 1 & 0\\
 -  \frac{s_{2\beta}v}{2 v_S} &   \frac{s_\beta s_{2\beta} v}{2 v_S} & 0 & 1\\
   c_\beta & s_\beta &\frac{s_{2\beta}v}{2 v_\chi}&   0
     \end{array}
     \right).
     \label{Rodd_sector}
\end{equation}

In general, the masses of the exotic scalar sector are proportional to 
$(a_{14} \tan\beta)^{1/2} v$, $v_{\chi}$ and $v_S$.


\section{Mass matrices}
\label{sect:Mass-Matrices}

\subsection{Up-like quark sector}

The mass matrix obtained from the Yukawa Lagrangian for the up-like quark sector in (\ref{eq:Lagrangian-quark-sector}) in the basis $(U^{1},U^{2},U^{3},T)$ is
\begin{equation}
\label{eq:Up-mass-matrix}
 M_{U}=\frac{1}{\sqrt{2}}
 \left(
 \begin{array}{cccc}
  0&(h_{2}^{U})_{12}v_{2}&0& (h_{2}^{T})_{1}v_{2}\\
  0&(h_{1}^{U})_{22}v_{1}&0& (h_{1}^{T})_{2}v_{1}\\
  (h_{1}^{U})_{31}v_{1}&0&(h_{1}^{U})_{33}v_{1}& 0\\
  0&(h_{\chi}^{U})_{2}v_{\chi}&0& h_{\chi}^{T}v_{\chi}
 \end{array}
 \right),
\end{equation}
Due to the hierarchy of the VEVs, the extended quadratic mass matrix for the up sector $M_{U}^{2}=M_{U} M_U^\dagger$ (see \ref{app:Fermion-mass-matrices} for the expression) can be written by blocks as follows
\begin{equation}
    M_{U}^{2}=
    \left(
    \begin{array}{cc}
    A&C\\
    C^{T}&D
    \end{array}
    \right),
\end{equation}
where $A\sim v_{1,2}^{2}$, $C\sim v_{1,2}v_{\chi}$ and $D\sim v_{\chi}^{2}$. Then, the matrix can be diagonalized by blocks
\begin{equation}
    \boldsymbol{m}_{U}^{2}=\left(V_{L}^{U}\right)^{T}M_{U}^{2}\left(V_{L}^{U}\right)=
    \left(
    \begin{array}{cc}
    m_{U}^{2} & 0\\
    0 & m_{T}^{2}
    \end{array}
    \right),
\end{equation}
where
\begin{equation}
    V_{L}^{U}=
    \left(
    \begin{array}{cccc}
    c_L^{u} & s_L^{u} & 0 & \frac{v_{2}}{v_{\chi}}\\
    -s_L^{u} & c_L^{u} & 0 & \frac{v_{1}}{v_{\chi}}\\
    0 & 0 & 1 & 0\\
    -\frac{v_{2}}{v_{\chi}} & -\frac{v_{1}}{v_{\chi}} & 0 & 1
    \end{array}
    \right).
\end{equation}
The block $D$ corresponds directly to the exotic $T$-quark mass.
After the diagonalization of the $M_{U}^{2}$ the eigenvalues for the up-sector are given by
\begin{align}
m_{u}^{2}&=0,\qquad
 m_{c}^{2}\approx \frac{1}{2}v_{1}^{2}\frac{[(h_{1}^{U})_{2}h_{\chi}^{T}-(h_{2}^{T})_{1}(h_{\chi}^{U})_{2}]^{2}}{((h_{\chi}^{U})_{2})^{2}+(h_{\chi}^{T})^{2}},\nonumber\\
 m_{t}^{2}&\approx \frac{1}{2}v_{1}^{2}\left[((h_{1}^{U})_{31})^{2}+((h_{1}^{U})_{33})^{2}\right],\nonumber\\
 m_{T}^{2}&\approx \frac{1}{2}v_{\chi}^{2}\left[((h_{\chi}^{U})_{2})^{2}+(h_{\chi}^{T})^{2}\right].
 \label{eq:upsectormass}
\end{align}
By checking the squared mass matrix $M_{U}^{2}$ (see (\ref{eq:mu2})), it is found that the charm quark $c$ plays a seesaw-like mechanism with the exotic $T$, as it is shown below in the submatrix of $2 M_{U}^{2}$, 
\begin{align*} 
\label{eq:charm-TTop}
\left(
\begin{array}{cc}
v_{1}^{2}[(h_{1}^{T})_{2}^{2}+(h_{1}^{U})_{22}^{2}] &
v_{1}v_{\chi}[(h_{1}^{T})_{2}(h_{\chi}^{T})+(h_{1}^{U})_{22}(h_{\chi}^{U})_{2}]\\
v_{1}v_{\chi}[(h_{1}^{T})_{2}(h_{\chi}^{T})+(h_{1}^{U})_{22}(h_{\chi}^{U})_{2}] &
v_{\chi}^{2}[(h_{\chi}^{T})^{2}+(h_{\chi}^{U})_{2}^{2}]
\end{array} 
\right). \nonumber
\end{align*}
Because of the hierarchy $v_{1}\ll v_{\chi}$, the smallest eigenvalue can be expressed as
\begin{equation*}
v_{1}^{2}[(h_{1}^{T})_{2}^{2}+(h_{1}^{U})_{22}^{2}] - 
\frac{(v_{1}v_{\chi}[(h_{1}^{T})_{2}(h_{\chi}^{T})+(h_{1}^{U})_{22}(h_{\chi}^{U})_{2}])^{2}}{v_{\chi}^{2}[(h_{\chi}^{T})^{2}+(h_{\chi}^{U})_{2}^{2}]}=
v_{1}^{2}\frac{[(h_{1}^{U})_{2}h_{\chi}^{T}-(h_{2}^{T})_{1}(h_{\chi}^{U})_{2}]^{2}}{((h_{\chi}^{U})_{2})^{2}+(h_{\chi}^{T})^{2}},
\end{equation*}
contains a subtraction of Yukawa couplings in the numerator of the expression for $m_{c}$, that allows a suppression of the associated numerical value from hundreds to units of GeV, since all the Yukawa couplings are assumed at the same order of magnitude. 

\subsection{Down-like quark sector}
For the down-like quark sector, the mass matrix in the basis $(D^{1},D^{2},D^{3},J^{1},J^{2})$ is given by
\begin{align}
\label{eq:Down-mass-matrix}
 M_{D}&=\frac{1}{\sqrt{2}}
 \left(
 \begin{array}{ccccc}
  0&0&0& (h_{1}^{J})_{11}v_{1}&(h_{1}^{J})_{12}v_{1}\\
  0&0&0& (h_{2}^{J})_{21}v_{2}&(h_{2}^{J})_{22}v_{2}\\
  0&0&(h_{2}^{D})_{33}v_{2}& 0&0\\
  0&0&0& (h_{\chi}^{J})_{11}v_{\chi}&0\\
  0&0&0& 0&(h_{\chi}^{J})_{22}v_{\chi}
 \end{array}
 \right).
\end{align}
The extended quadratic mass matrix for the down sector $M_{D}^{2}=M_{D} M_D^\dagger$ (see \ref{app:Fermion-mass-matrices} for the expression) can be written by blocks
\begin{equation}
    M_{D}^{2}=
    \left(
    \begin{array}{cc}
    A & C\\
    C^{T} & D
    \end{array}
    \right),
\end{equation}
where $A\sim v_{1,2}^{2}$, $C\sim v_{1,2}v_{\chi}$ and $D\sim v_{\chi}^{2}$. After block diagonalization, the matrix becomes
\begin{equation}
    \boldsymbol{m}_{D}^{2}=\left(V_{L}^{D}\right)^{T}M_{D}^{2}\left(V_{D}^{U}\right)=
    \left(
    \begin{array}{cc}
    m_{D}^{2} & 0\\
    0 & m_{J}^{2}
    \end{array}
    \right),
\end{equation}
where $m_{J}^{2}\sim D$ and $m_{D}^{2}$ corresponds to the SM sector, given the following eigenvalues
\begin{align}
 m_{d}&=0,
 &m_{s}&=0,\nonumber\\
 m_{b}&=\frac{1}{\sqrt{2}}(h_{2}^{D})_{33}v_{2},
 &m_{J}^{i}&=\frac{1}{\sqrt{2}}(h_{\chi}^{J})_{ii}v_{\chi}.
\label{eq:Bottom}
\end{align}

\begin{figure}[h]
\centering
\includegraphics[scale=0.7]{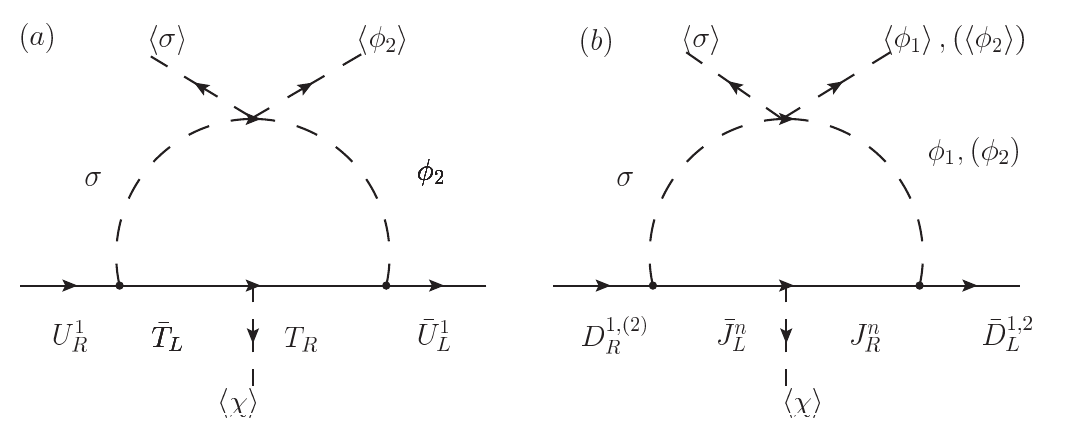}
\caption{One-loop corrections to the self-energies in the (a) up and (b) down quark sectors.}
\label{fig:Radiative Corrections}
\end{figure}

\subsubsection{Radiative corrections.}

The $u,d$ and $s$ quarks turn out to be massless, so it is necessary to introduce radiative corrections to the model. These corrections are shown in the Figure (\ref{fig:Radiative Corrections}a),  for the up sector, which add small contributions to the $M_U$ matrix with the term
\begin{equation}
\label{eq:Rad-Corr-Up}
\Sigma _{11}^{(u)}=\frac{\lambda_{7}^{\prime}\left<\sigma\right>v_{2}\left(h^U_{\sigma}\right)_1\left(h^T_2\right)_1}{\sqrt{2}M_T}C_0\left(\frac{M_2}{m_T},\frac{m_{\sigma}}{m_T}\right),
\end{equation} 
where
\begin{equation}
C_0\left(x_1,x_2\right)=\frac{1}{16\pi ^2}\frac{1}{\left(1-x_1^2\right)\left(1-x_2^2\right)\left(x_1^2-x_2^2\right)}
\left[x_1^2x_2^2\ln\left(\frac{{x_2^2}}{x_1^2}\right)+x_1^2\ln x_1^2-x_2^2 \ln x_2^2\right],
\label{oneloop-coef}
\end{equation}
and $\left<\sigma\right>$ is the VEV of the $\sigma$ scalar field. In this case the extended $M_{U}^{2}$ matrix (\ref{eq:mu2}) is corrected with new contributions as
\begin{equation}
    M_{U}^{2}=
    \left(
    \begin{array}{cccc}
    \Sigma_{11}^{(u)} & 0 & \Sigma_{13}^{(u)} & 0\\
    \Sigma_{21}^{(u)} &m_{c} & \Sigma_{23}^{(u)} &0\\
    0 & 0 & m_{t}& 0\\
    0 & 0 & 0 &m_{T}
    \end{array}
    \right)
    \left(
     \begin{array}{cccc}
    \Sigma_{11}^{(u)} & \Sigma_{21}^{(u)} & 0 & 0\\
    0 &m_{c} & 0 &0\\
    \Sigma_{13}^{(u)} & \Sigma_{23}^{(u)} & m_{t}& 0\\
    0 & 0 & 0 &m_{T}
    \end{array}
    \right),
\end{equation}
where the matrix associated to the SM sector can be diagonalized through the new corrected matrix
\begin{equation}
    V_{L_{SM}}^{U}=
    \left(
    \begin{array}{ccc}
    c_L^{u} & s_L^{u} &\Sigma_{13}^{(u)}/m_{t}\\
    -s_L^{u} & c_L^{u} &\Sigma_{23}^{(u)}/m_{t}\\
    -\Sigma_{13}^{(u)}/m_{t} & -\Sigma_{23}^{(u)}/m_{t} & 1
    \end{array}
    \right).
\end{equation}
For the down sector, taking into account the coupling with the $\sigma$ scalar field showed in diagram (b), the self-energies generated at one-loop level for down and strange quarks have the form
\begin{align}
\label{eq:Rad-Corr-Down}
\Sigma _{1a}^{(d)}&=\sum_{n=1,2}\frac{\lambda_{6}^{\prime}\left<\sigma\right>v_{1}\left(h^{J}_{1}\right)_{1n}\left(h^D_{\sigma }\right)_{na}}{\sqrt{2}M_{J^{n}}}C_0\left(\frac{M_{1}}{M_{J^{n}}},\frac{M_{\sigma}}{M_{J^{n}}}\right),\\
\label{eq:Rad-Corr-Strange}
\Sigma _{2a}^{(d)}&=\sum_{n=1,2}\frac{\lambda_{7}^{\prime}\left<\sigma\right>v_{2}\left(h^{J}_{2}\right)_{2n}\left(h^D_{\sigma }\right)_{na}}{\sqrt{2}M_{J^{n}}}C_0\left(\frac{M_{2}}{M_{J^{n}}},\frac{M_{\sigma}}{M_{J^{n}}}\right),
\end{align}
with $a=1,2$.

The previous article \cite{Mantilla} found that the self-energies $\Sigma_{13}$, $\Sigma_{23}$, $\Sigma_{31}$ and $\Sigma_{32}$ are different from zero because the three $D_R^j$ were universal under $X$ charge and the Lagrangian needed to generate these self-energies was
\begin{equation}
\overline{q_L^{1}} (\phi_{1}h^{J}_{1})_{1n} J^{n}_{R}+\overline{q_L^{a}}\left(\phi_{2}h^{J}_{2}\right)_{an} J^{n}_{R}
+\overline{J_{L}^n}\left(\sigma^{*}h_{\sigma }^{D}\right)_{nj}{D}_{R}^{j}+h.c.,
 \label{eq:Lagrangian-downquarkS-sector}
\end{equation}
where $j=1,2,3$ and $a=2,3$. However, in the present model, the PQ charges are not universal for the right-handed down quarks, and the Lagrangian shown above does not allow the term $j=3$ and then the self-energies $\Sigma_{13}$, $\Sigma_{23}$, $\Sigma_{31}$ and  $\Sigma_{32}$ vanish. Consequently, the mass matrix for the SM down quark sector can be written as
\begin{equation}
M_D=
\left(
    \begin{array}{ccc}
   \Sigma_{11}^{(d)} & \Sigma_{12}^{(d)} & 0\\
   \Sigma_{12}^{(d)} & \Sigma_{22}^{(d)} & 0 \\
   0 & 0 & m_b
     \end{array}
     \right).
     \label{M_D}
\end{equation}
The bottom quark is now decoupled from the lightest quarks and the rotation matrices that diagonalize the mass matrices $M_D M_D^{\dagger}$ and $M_D^{\dagger} M_D$ are
\begin{equation}
V_{L,R}^{(d)}=
\left(
    \begin{array}{ccc}
   \cos\theta_{L,R}^{(d)} & \sin\theta_{L,R}^{(d)} & 0\\
   -\sin\theta_{L,R}^{(d)} & \cos\theta_{L,R}^{(d)} & 0 \\
   0 & 0 & 1
     \end{array}
     \right).
     \label{VLR}
\end{equation}
The expressions for $\tan\theta_L^{(d)}$ and $\tan\theta_R^{(d)}$ 
are given by
\begin{align}
\tan 2\theta_L^{(d)} &=\frac{2(\Sigma_{11}^{(d)}\Sigma_{21}^{(d)}+\Sigma_{12}^{(d)}\Sigma_{22}^{(d)})}{(\Sigma_{22}^{(d)})^2-(\Sigma_{11}^{(d)})^2+(\Sigma_{21}^{(d)})^2-(\Sigma_{12}^{(d)})^2}, \nonumber \\
\tan 2\theta_R^{(d)} &=\frac{2(\Sigma_{11}^{(d)}\Sigma_{12}^{(d)}+\Sigma_{21}^{(d)}\Sigma_{22}^{(d)})}{(\Sigma_{22}^{(d)})^2-(\Sigma_{11}^{(d)})^2-(\Sigma_{21}^{(d)})^2+(\Sigma_{12}^{(d)})^2}, 
\end{align}
and, after replacing the expressions for the self-energies, we find
\begin{align}
&\tan\theta_L^{(d)} = \frac{2 v_1 v_2 \lambda_6^{\prime}\lambda_7^{\prime}(h_1^J)_{11}(h_2^J)_{21}}{\lambda^{\prime 2}_7 v_2^2(h_2^J)_{21}^2-\lambda^{\prime 2}_6 v_1^2(h_1^J)_{11}^2},\nonumber  \\ 
&\tan\theta_R^{(d)} =\frac{2 v_1 v_2 \lambda_6^{\prime}\lambda_7^{\prime}(h_1^J)_{11}(h_2^J)_{21}}{\lambda^{\prime 2}_7 v_2^2(h_2^J)_{21}^2-\lambda^{\prime 2}_6 v_1^2(h_1^J)_{11}^2} \frac{(h_\sigma^D)_{11}^2+(h_\sigma^D)_{12}^2}{(h_\sigma^D)_{12}^2-(h_\sigma^D)_{11}^2}.
\end{align}
 In the limit where $v_1\approx m_t\gg v_2\approx m_b$ and considering that $(h_\sigma^D)_{12}$ and $(h_\sigma^D)_{11}^2$ are of the same order, we have for the left and right mixing angles
\begin{align}
\theta_L^{(d)} &\approx - \frac{ \lambda_6' v_2}{\lambda_7' v_1},\nonumber \\
\theta_R^{(d)} &\approx \frac{\pi}{4}.
\end{align} 
with the bottom quark completely decoupled from the down and strange quarks. 

\subsubsection{Numerical exploration for the quark sector.}
\label{eq:Num-Quarks}
The parameter space of the masses of the SM quarks was explored using a Montecarlo procedure to find the allowed regions consistent to the current experimental values reported in \cite{PDG}. It was assumed an electroweak VEV of $v=246.22 \mathrm{\,GeV}$ and $\tan\beta=175$. The main constrains of the exploration consists on (1) setting that any Higgs or Yukawa coupling spans the interval $[0.1,4\pi]$ to ensure naturalness and perturbativity, (2) the masses of the exotic fermions are set to the interval $[1\mathrm{\,TeV},5\mathrm{\,TeV}]$, (3) the mass of the singlet $m_{\sigma}$ spans the interval $[300\mathrm{\,GeV},900\mathrm{\,GeV}]$ and $\left< \sigma \right>\in [0.1\mathrm{\,TeV},1\mathrm{\,TeV}]$. 

\begin{figure}
\centering 
\subfloat[Up quark $u$.]{
	\begin{minipage}[c][.3\textwidth]{
	   0.30\textwidth}
	   \centering
	   \includegraphics[scale=0.50]{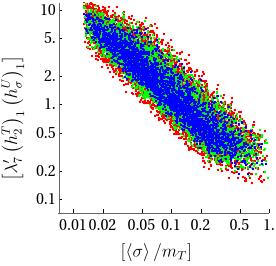}
	\end{minipage}}
\subfloat[Down quark $d$.]{
	\begin{minipage}[c][.3\textwidth]{
	   0.30\textwidth}
	   \centering
	   \includegraphics[scale=0.50]{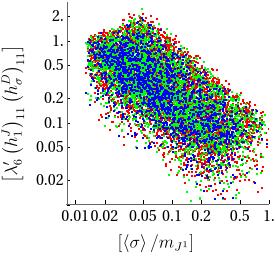}
	\end{minipage}}
\subfloat[Strange quark $s$.]{
	\begin{minipage}[c][.3\textwidth]{
	   0.30\textwidth}
	   \centering
	   \includegraphics[scale=0.50]{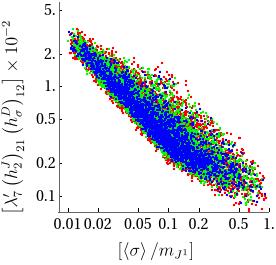}
	\end{minipage}}

\subfloat[Charm quark $c$.]{
	\begin{minipage}[c][.3\textwidth]{
	   0.32\textwidth}
	   \centering
	   \includegraphics[scale=0.50]{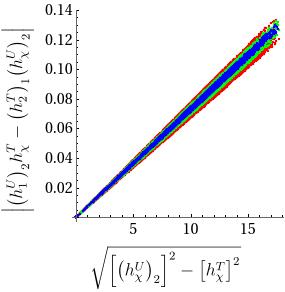}
	\end{minipage}}
\subfloat[Top quark $t$.]{
	\begin{minipage}[c][.3\textwidth]{
	   0.32\textwidth}
	   \centering
	   \includegraphics[scale=0.50]{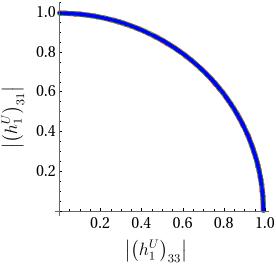}
	\end{minipage}}
\caption{Allowed regions for the parameter space consistent at $1\sigma$, $2\sigma$ and $3\sigma$.}
\label{fig:Quarks-num}
\end{figure}

The mass of the bottom quark $b$ in Eq. \eqref{eq:Bottom} does not require to be numerically explored since its expression allows a direct comparison to the estimated experimental mass $m_{b}=4.18^{+0.03}_{-0.02}\mathrm{\,GeV}$ reported also in \cite{PDG} yielding:
\begin{equation}
\label{eq:Bottom-num}
(h_{2}^{D})_{33} = 4.24^{+0.04}_{-0.03} = 
4\pi\left(0.334^{+0.003}_{-0.002} \right).
\end{equation}
On the other hand, the tree-level masses of the charm $c$ and top $t$ quarks in Eq. \eqref{eq:upsectormass} were explored numerically to show the consistency of the model with the experimental values $m_{c}=1.27\pm 0.02\mathrm{\,GeV}$ and $m_{t}=172.76\pm0.30\mathrm{\,GeV}$ at $1\sigma$, $2\sigma$ and $3\sigma$ according to \cite{PDG}. The main results are shown in the Figures \ref{fig:Quarks-num}(d) and \ref{fig:Quarks-num}(e). 

Regarding the up quark $u$, the radiative correction $\Sigma _{11}^{(u)}$ in Eq. \eqref{eq:Rad-Corr-Up} was explored to find the parameter space consistent with
\begin{equation}
\label{eq:Up-exp}
\Sigma _{11}^{(u)} = 2.16^{+0.49}_{-0.26}\mathrm{\,MeV},
\end{equation}
at $1\sigma$, $2\sigma$ and $3\sigma$. The CP-even mass eigenstates associated to the scalar boson $\phi_{2}$ running in the loop in Figure \ref{fig:Radiative Corrections}(a) are substituted in \eqref{eq:Rad-Corr-Up} obtaining the individual contributions of $h_{\mathrm{SM}}$ and $H$ as
\begin{equation}
\Sigma _{11}^{(u)}=
\Sigma_{11}^{(uh_{\mathrm{SM}})}
+\Sigma _{11}^{(uH)},
\end{equation} 
where the main contribution is done by
\begin{equation}
\Sigma_{11}^{(uh_{\mathrm{SM}})} =
\frac{c_{\beta}s_{\beta}^{2} v }{\sqrt{2}}
\frac{\lambda_{7}^{\prime}
\left(h^U_{\sigma}\right)_1 \!\!
\left(h^T_2\right)_1		\!\!
\left<\sigma\right>}{m_T} 
C_0\left(\frac{m_{h_{\mathrm{SM}}}}{m_{T}},\frac{m_{\sigma}}{m_{T}}\right),
\end{equation} 
and the term $\Sigma_{11}^{(uH)} $ is strongly suppressed by the factor $c_{\beta}^{3}$, 
\begin{equation}
\Sigma_{11}^{(uH)} =
\frac{c_{\beta}^{3} v }{\sqrt{2}}
\frac{\lambda_{7}^{\prime}
\left(h^U_{\sigma}\right)_1	\!\!
\left(h^T_2\right)_1		\!\!
\left<\sigma\right>}{m_T} 
C_0\left(\frac{m_{H}}{m_{T}},\frac{m_{\sigma}}{m_{T}}\right).
\end{equation} 
The parameter space $\lambda_{7}^{\prime}
\left(h^U_{\sigma}\right)_1	\!\!
\left(h^T_2\right)_1$ vs. $\left<\sigma\right>\!/m_{T}$ consistent to Eq. \eqref{eq:Up-exp} is shown in the Figure \ref{fig:Quarks-num}(a). 

Similarly, the radiative correction $\Sigma _{11}^{(d)}$ in Eq. \eqref{eq:Rad-Corr-Down} for the down quark $d$ was also explored to find the allowed region consistent with
\begin{equation}
\label{eq:Down-exp}
\Sigma _{11}^{(d)} = 4.67^{+0.48}_{-0.17}\mathrm{\,MeV}. 
\end{equation}
Although there are contributions from the two exotic species $J^{1}$ and $J^{2}$, it would be the lightest one, named $J^{1}$, which does the main contribution to the self-energy. On the other hand, the mass eigenstates of the CP-even scalars associated to $\phi_{1}$ running in the loop in Figure \ref{fig:Radiative Corrections}(b) make two contributions:
\begin{align}
\Sigma _{11}^{(d)}&=
\frac{\lambda_{6}^{\prime}
\left<\sigma\right>\!v_{1}\!
\left(h^{J}_{1}\right)_{11}\!
\left(h^D_{\sigma }\right)_{11}
}{\sqrt{2}m_{J^{1}}}
C_0\left(
\frac{M_{1}}{m_{J^{1}}},
\frac{m_{\sigma}}{m_{J^{1}}}\right),
\end{align}
the individual contributions of $h_{\mathrm{SM}}$ and $H$ as
\begin{equation}
\Sigma _{11}^{(d)}=
\Sigma_{11}^{(dh_{\mathrm{SM}})}
+\Sigma _{11}^{(dH)},
\end{equation} 
where each of the contributions are
\begin{equation}
\Sigma_{11}^{(dh_{\mathrm{SM}})} = 
\frac{c_{\beta}^{2}s_{\beta}v }{\sqrt{2}}
\frac{\lambda_{6}^{\prime}\!
\left(h^{J}_{1}\right)_{11}\!
\left(h^D_{\sigma }\right)_{11}\!
\left<\sigma\right>}{m_{J^{1}}} 
C_0\left(
\frac{m_{h_{\mathrm{SM}}}}{m_{J^{1}}},
\frac{m_{\sigma}}{m_{J^{1}}}\right),
\end{equation} 
and
\begin{equation}
\Sigma_{11}^{(dH)} =
\frac{s_{\beta}^{3}v }{\sqrt{2}}
\frac{\lambda_{6}^{\prime}\!
\left(h^{J}_{1}\right)_{11}\!
\left(h^D_{\sigma }\right)_{11}\!
\left<\sigma\right>}{m_{J^{1}}} 
C_0\left(\frac{m_{H}}{m_{J^{1}}},
\frac{m_{\sigma}}{m_{J^{1}}}\right). 
\end{equation} 
The parameter space $\lambda_{6}^{\prime}\!
\left(h^{J}_{1}\right)_{11}\!
\left(h^D_{\sigma }\right)_{11}$ vs. $\left<\sigma\right>\!/m_{J^{1}}$ consistent to Eq. \eqref{eq:Down-exp} is shown in the Figure \ref{fig:Quarks-num}(b). 

Lastly, the self-energy that contributes to the mass of the strange quark $s$ is explored such that, at $1\sigma$, $2\sigma$ and $3\sigma$, it fulfills
\begin{equation}
\label{eq:Strange-exp}
\Sigma _{22}^{(d)} = 93^{+11}_{-05}\mathrm{\,MeV}. 
\end{equation}
In the same way as the down quark $d$, only the contribution of the lightest exotic quark $J^{1}$ is assumed as the main contribution to the self-energy. The contributions done by the mass eigenstates of the CP-even scalars associated to $\phi_{2}$ are
\begin{align}
\Sigma _{22}^{(d)}&=
\frac{\lambda_{7}^{\prime}
\left<\sigma\right>\!v_{2}\!
\left(h^{J}_{2}\right)_{21}\!
\left(h^D_{\sigma }\right)_{12}
}{\sqrt{2}m_{J^{1}}}
C_0\left(
\frac{M_{1}}{m_{J^{1}}},
\frac{m_{\sigma}}{m_{J^{1}}}\right),
\end{align}
the individual contributions of $h_{\mathrm{SM}}$ and $H$ as
\begin{equation}
\Sigma _{22}^{(d)}=
\Sigma_{22}^{(dh_{\mathrm{SM}})}
+\Sigma _{22}^{(dH)},
\end{equation} 
where each of the contributions are
\begin{equation}
\Sigma_{22}^{(dh_{\mathrm{SM}})} = 
\frac{c_{\beta}s_{\beta}^{2} v }{\sqrt{2}}
\frac{\lambda_{7}^{\prime}
\left(h^{J}_{2}\right)_{21}\!
\left(h^D_{\sigma }\right)_{12}
\left<\sigma\right>}{m_{J^{1}}}
C_0\left(
\frac{m_{h_{\mathrm{SM}}}}{m_{J^{1}}},
\frac{m_{\sigma}}{m_{J^{1}}}\right),
\end{equation} 
and
\begin{equation}
\Sigma_{22}^{(uH)} =
\frac{c_{\beta}^{3}v }{\sqrt{2}}
\frac{\lambda_{7}^{\prime}
\left(h^{J}_{2}\right)_{21}\!
\left(h^D_{\sigma }\right)_{12}
\left<\sigma\right>}{m_{J^{1}}} 
C_0\left(\frac{m_{H}}{m_{J^{1}}},
\frac{m_{\sigma}}{m_{J^{1}}}\right). 
\end{equation} 
The parameter space $\lambda_{7}^{\prime}
\left(h^{J}_{2}\right)_{21}\!
\left(h^D_{\sigma }\right)_{12}$ vs. $\left<\sigma\right>\!/m_{J^{1}}$ consistent to Eq. \eqref{eq:Down-exp} is shown in the Figure \ref{fig:Quarks-num}(c). 

The masses of the exotic quarks $T$ and $J^a$ are proportional to $v_\chi$, so it is possible to take them at order of TeV according to the LHC bounds (see Sec. 2.1). On the other hand, the Yukawa couplings can be considered as free parameters.

\subsection{Charged lepton sector}
\begin{figure}
\centering
\includegraphics[scale=0.6]{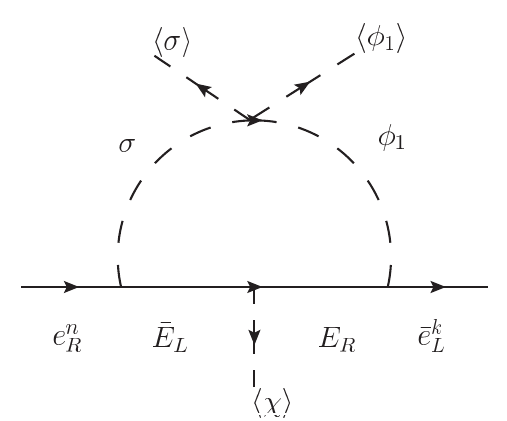}
\caption{One-loop correction to the self-energy of the electron $e$.}
\label{fig:RC-electron}
\end{figure}

The Yukawa Lagrangian for the lepton sector in (\ref{eq:Leptonic-Lagrangian}) yields the following mass matrix for the charged leptons in the basis $(e^{e},e^{\mu},e^{\tau},E,\mathcal{E})$
\begin{equation}
\label{eq:Electron-mass-matrix}
   M_{E}=\frac{1}{\sqrt{2}}
   \left( \begin{array}{ccccc}
        0&g^{2e}_{e\mu}v_{2}&0 &g^{1}_{Ee}v_{1}&0\\
        0&g^{2e}_{\mu\mu}v_{2}&0 &g^{1}_{E\mu}v_{1}&0\\
        g^{2e}_{\tau e}v_{2}&0&g^{2e}_{\tau\tau}v_{2} &0&0\\
       
        0&0&0& h^{\chi E}v_{\chi}&0\\
        0&0&0& 0&h^{\chi\mathcal{E}}v_{\chi}
    \end{array}
    \right).
\end{equation}
Similarly to the quarks of the first generation, after diagonalizing $M_{E}^{2}=M_{E} M_E^\dagger$ \eqref{eq:me2}, the electron does not acquire mass at tree level, so it is necessary to implement radiative corrections in order to produce a finite mass. The couplings to the $\sigma$-singlet yield the self-energies depicted in the figure \ref{fig:RC-electron}, and expressed by
\begin{align}
\label{eq:Rad-Corr-Electron}
 \Sigma_{11(13)}^{(e)}&=\frac{\lambda_{6}^{\prime}\left<\sigma\right>v_{1}(g_{Ee}^{1})(h_{E}^{\sigma e(\tau)})}{m_{E}}C_{0}\left(\frac{M_{1}}{m_{E}},\frac{m_{\sigma}}{m_{E}}\right), \nonumber \\
\Sigma_{21(23)}^{(e)}&=\frac{\lambda_{6}^{\prime}\left<\sigma\right>v_{1}(g_{E\mu}^{1})(h_{E}^{\sigma e(\tau)})}{m_{E}}C_{0}\left(\frac{M_{1}}{m_{E}},\frac{m_{\sigma}}{m_{E}}\right).
\end{align}
which contribute to the mass matrix in the form
\begin{equation}
   \Delta M_{E}=
   \left( \begin{array}{ccc}
        \Sigma_{11}^{(e)}&0&\Sigma_{13}^{(e)} \\
        \Sigma_{21}^{(e)}&0&\Sigma_{23}^{(e)}\\
        0&0&0
    \end{array}
    \right).
\end{equation}
The associated mass eigenvalues to the $M_{E}+\Delta M_{E}$ matrix are
\begin{align}
\label{eq:Charged-Leptons-masses}
    m_{e}&\approx\Sigma_{11}^{(e)},
    &m_{\mu}&=\frac{v_{2}}{\sqrt{2}}\left[\left(g_{e\mu}^{2e}\right)^{2}+\left(g_{\mu\mu}^{2e}\right)^{2}\right]^{1/2},\nonumber\\
    m_{\tau}&=\frac{v_{2}}{\sqrt{2}}\left[\left(g_{\tau  e}^{2e}\right)^{2}+\left(g_{\tau\tau}^{2e}\right)^{2}\right]^{1/2},
    &m_{E}&=\left(h^{\chi E}\right)\frac{v_{\chi}}{\sqrt{2}},\\ m_{\mathcal{E}}&=\left(h^{\chi \mathcal{E}}\right)\frac{v_{\chi}}{\sqrt{2}}.\nonumber
\end{align}
And the matrix rotations of the  $3\times 3$ submatrices of the $M_{E} M_{E}^{\dagger}$ and  $M_{E}^{\dagger} M_{E}$ are given respectively by
\begin{align}
V_{L}^{(e)}&=
\left(
    \begin{array}{ccc}
   \cos\alpha_{L}^{(e)} & \sin\alpha_{L}^{(e)} & \frac{\Sigma_{13}}{m_\tau}\\
   -\sin\alpha_{L}^{(e)} & \cos\alpha_{L}^{(e)} & \frac{\Sigma_{23}}{m_\tau} \\
    -\frac{\Sigma_{13}}{m_\tau} & -\frac{\Sigma_{23}}{m_\tau} & 1
     \end{array}
     \right),\nonumber\\
V_{R}^{(e)}&=
\left(
    \begin{array}{ccc} 
    \cos\alpha_{R}^{(e)} &  \frac{\Sigma_{12}}{m_\mu}  &\sin\alpha_{R}^{(e)} \\
    \frac{\Sigma_{21}}{m_\mu} & 1 & \frac{m_\mu\Sigma_{23}}{m_\tau^2} \\
   -\sin\alpha_{R}^{(e)} &   -\frac{m_\mu\Sigma_{23}}{m_\tau^2}  & \cos\alpha_{R}^{(e)} 
     \end{array}
     \right).     
     \label{VLRe}
\end{align}
where the angles are given by the expressions
\begin{equation}
\tan\alpha_L^{(e)} = \frac{g_{e\mu}^{2e}}{g_{\mu\mu}^{2e}}, \;\;\;
\tan\alpha_R^{(e)}=\frac{g_{\tau e}^{2e}}{g_{\tau\tau}^{2e}}.
\end{equation}
The case where $g_{e\mu}^{2e}\ll g_{\mu\mu}^{2e}$ and $g_{\tau\tau}^{2e} \approx g_{\tau e}^{2e}$ can be considered without producing problems in the mass hierarchy of the charged leptons. In this case $\tan\alpha_L^{(e)}\approx \sin\alpha_L^{(e)}$ and  $\alpha_R^{(e)} =\pi/4$. On the other hand, the hierarchies in the Yukawa couplings $h_E^{\sigma e}/h_E^{\sigma\tau}\approx m_{e}/m_{\mu}$ and $g_{Ee}^1 \approx g_{E\mu}^1 $ are considered, allowing to approximate the self-energies $\Sigma^{e}_{13} \approx m_e$ and $\Sigma^{e}_{23} \approx m_{\mu}$. With this approximation is useful to estimate the order of the matrix inputs (which are taken into account to study flavour changing with the axion field) as
\begin{align} 
V_{L}^{(e)}&\approx
\left(
    \begin{array}{ccc}
   1 & s_{\alpha_{L}}^{(e)} & \frac{m_e}{m_\tau}\\
   -s_{\alpha_{L}}^{(e)} & 1 & \frac{m_\mu}{m_\tau} \\
   - \frac{m_e}{m_\tau} & - \frac{m_\mu}{m_\tau} & 1
     \end{array}
     \right),\nonumber\\
V_{R}^{(e)}&\approx
\left(
    \begin{array}{ccc} 
    \frac{1}{\sqrt{2}}&  \frac{m_e}{m_\mu}  &\frac{1}{\sqrt{2}} \\
    -\frac{m_e}{m_\mu} & 1 & 0\\
   - \frac{1}{\sqrt{2}} &   0  & \frac{1}{\sqrt{2}}  \\  
     \end{array}
     \right).     
     \label{AVLR}
\end{align}

Finally, after diagonalizing the charged leptonic mass matrix and using (\ref{eq:Leptonic-Lagrangian}), the couplings with the Higgs $h_{SM}$ turn out to be the same of the SM.

\subsubsection{Numerical exploration for the charged lepton sector.}
\label{eq:Num-Leptons}
The parameter space of the masses of the SM charged leptons was also explored using a Montecarlo procedure to find the allowed regions consistent with the current experimental values reported in \cite{PDG}. The assumptions of an electroweak VEV of $v=246.22 \mathrm{\,GeV}$ and $\tan\beta=175$, as well as the intervals for the dimensionless couplings and the masses of exotic species reported in \ss \ref{eq:Num-Quarks} are hold.

\begin{figure}
\centering 
\subfloat[Electron $e$.]{
	\begin{minipage}[c][.3\textwidth]{
	   0.30\textwidth}
	   \centering
	   \includegraphics[scale=0.50]{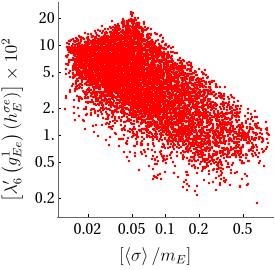}
	\end{minipage}}
\subfloat[Muon $\mu$.]{
	\begin{minipage}[c][.3\textwidth]{
	   0.30\textwidth}
	   \centering
	   \includegraphics[scale=0.50]{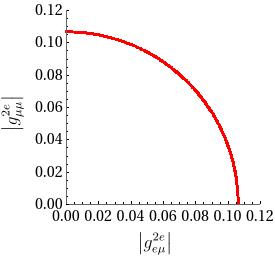}
	\end{minipage}}
\subfloat[Tau $\tau$.]{
	\begin{minipage}[c][.3\textwidth]{
	   0.30\textwidth}
	   \centering
	   \includegraphics[scale=0.20]{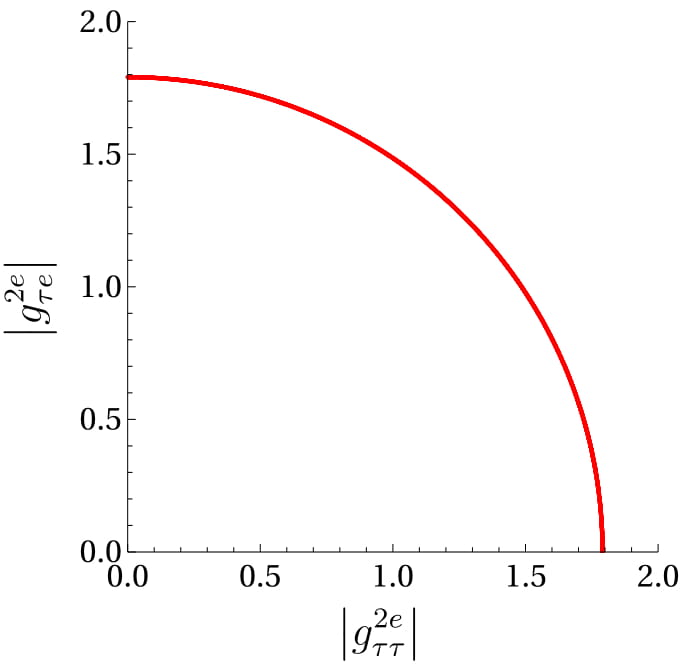}
	\end{minipage}}
\caption{Allowed regions for the parameter space consistent to the masses reported in \cite{PDG}.}
\label{fig:Leptons-num}
\end{figure}

The masses of the two heaviest charged leptons, $\mu$ and $\tau$, in Eq. \eqref{eq:Charged-Leptons-masses} were explored numerically to show the consistency of the model with the experimental values $m_{\mu}=105.7\mathrm{\,MeV}$ and $m_{\tau}=1776.86\pm0.12\mathrm{\,MeV}$ at $1\sigma$, $2\sigma$ and $3\sigma$ according to \cite{PDG}. In the case of the muon, because of its narrow experimental uncertainty, a width of $0.1\mathrm{\,MeV}$ was assumed to make the procedure able to find solutions. The main results are shown in the Figures \ref{fig:Leptons-num}(b) and \ref{fig:Leptons-num}(c). 
It is worth noting that the mass of the muon is one order of magnitude smaller than the mass of the tau. The ratio of the masses is given by 
\begin{equation}
\frac{m_{\tau}}{m_{\mu}} = 
\sqrt{
\frac{\left(g_{\tau  e}^{2e}\right)^{2}+\left(g_{\tau\tau}^{2e}\right)^{2}}
{\left(g_{e\mu}^{2e}\right)^{2}+\left(g_{\mu\mu}^{2e}\right)^{2}}
} \approx 17.
\end{equation}
Two interesting limits consist on taking the cases when the two couplings are equal, $g_{e\mu}^{2e}\approx g_{\mu\mu}^{2e}\approx 0.08$ and $g_{\tau  e}^{2e}\approx g_{\tau\tau}^{2e}\approx 1.26$, or when one coupling dominates, $g_{e\mu}^{2e}\ll g_{\mu\mu}^{2e}\approx 0.10$ and $g_{\tau  e}^{2e}\approx g_{\tau\tau}^{2e}\approx 1.78$.
Despite the fact that this ratio is of the order of 17, the Yukawa couplings that generate the masses of the muon and tau are perturbative.

Regarding to the mass of the electron $m_e$ which is approximately determined by the self-energy $\Sigma _{11}^{(e)}$ in Eq. \eqref{eq:Rad-Corr-Electron} was explored to find the parameter space consistent with
\begin{equation}
\label{eq:Electron-exp}
\Sigma _{11}^{(e)} = 0.511\mathrm{\,MeV},
\end{equation}
with a width of $0.05\mathrm{\,MeV}$ to be able to find solutions. Similarly to the down quark $d$, the two contributions coming from the CP-even mass eigenstates $h_{\mathrm{SM}}$ and $H$ from the $\phi_{1}$ running in the loop in Figure \ref{fig:RC-electron} are
\begin{equation}
\Sigma _{11}^{(e)}=
\Sigma_{11}^{(eh_{\mathrm{SM}})}
+\Sigma _{11}^{(eH)},
\end{equation} 
where each one is give by
\begin{equation}
\Sigma_{11}^{(eh_{\mathrm{SM}})} =
\frac{c_{\beta}^{2}s_{\beta}v }{\sqrt{2}}
\frac{\lambda_{6}^{\prime}
\left(g^{1}_{Ee}\right)	\!
\left(h^{\sigma e}_{E}\right)	\!
\left<\sigma\right>}{m_E} 
C_0\left(
\frac{m_{h_{\mathrm{SM}}}}{m_{E}},
\frac{m_{\sigma}}{m_{E}}\right),
\end{equation} 
and
\begin{equation}
\Sigma_{11}^{(eH)} =
\frac{s_{\beta}^{3}v }{\sqrt{2}}
\frac{\lambda_{6}^{\prime}
\left(g^{1}_{Ee}\right)	\!
\left(h^{\sigma e}_{E}\right)	\!
\left<\sigma\right>}{m_E} 
C_0\left(
\frac{m_{H}}{m_{E}},
\frac{m_{\sigma}}{m_{E}}\right).
\end{equation} 
The parameter space $\lambda_{6}^{\prime}
\left(g^{1}_{Ee}\right)	\!
\left(h^{\sigma e}_{E}\right)$ vs. $\left<\sigma\right>\!/m_{E}$ consistent to Eq. \eqref{eq:Up-exp} is shown in the Figure \ref{fig:Leptons-num}(a). 

For the exotic leptons we take into account the same comments for the exotic quarks.  These masses are also proportional to the $v_\chi$. So, they are bigger than $1$ TeV and your Yukawa parameters can be taken as free parameters.

\subsection{Neutral lepton sector}

Regarding the neutrino sector, the Yukawa Lagrangian is given by
\begin{equation}
    -\mathcal{L}=h_{2e}^{\nu i}\bar{\ell}_{L}^{e}\tilde{\phi}_{2}\nu_{R}^{i}+h_{2\mu}^{\nu i}\bar{\ell}_{L}^{\mu}\tilde{\phi}_{2}\nu_{R}^{i}+h^{ij}_{S}\bar{\nu}_{R}^{C i}S\nu_{R}^{j},
\end{equation}
where $i,j=e,\mu,\tau$. After the SSB, the mass Lagrangian has the structure
\begin{equation}
    -\mathcal{L}_{\text{mass}}=h_{2e}^{\nu i}\frac{v_{2}}{\sqrt{2}}\bar{\nu}_{L}^{e}\nu_{R}^{i}
    +h_{2\mu}^{\nu i}\frac{v_{2}}{\sqrt{2}}\bar{\nu}_{L}^{\mu}\nu_{R}^{i}
    +h_{S}^{ij}\frac{v_{S}}{\sqrt{2}}\bar{\nu}_{L}^{C i}\nu_{R}^{j}.
\end{equation}
Without loss of generality, we consider $h_{S}^{ij}$ as a diagonal matrix, where
\begin{equation}
    \frac{h_{S}^{ij}v_{S}}{\sqrt{2}}=M^{i}\delta_{i}^{j},
\end{equation}
are the masses of the three right-handed neutrinos. Thus, in the $\left(\nu_{L},\nu_{R}^{C}\right)$ basis the mass matrix can be written down as
\begin{equation}
    M_{\nu}=
    \left(
    \begin{array}{cc}
    0&m_{D}^{T}\\
    m_{D}&M_{M}
    \end{array}
    \right),
\end{equation}
where
\begin{align}
    m_{D}&=\frac{v_{2}}{\sqrt{2}}
    \left(
    \begin{array}{ccc}
    h_{2e}^{\nu e}&h_{2e}^{\nu\mu}&h_{2e}^{\nu\tau}\\
    h_{2\mu}^{\nu e}&h_{2\mu}^{\nu\mu}&h_{2\mu}^{\nu\tau}\\
    0&0&0
    \end{array}
    \right),\\
    M_{M}&=\frac{v_{S}}{\sqrt{2}}
    \left(
    \begin{array}{ccc}
    h_{1}&0&0\\
    0&h_{2}&0\\
    0&0&h_{3}
    \end{array}
    \right).
\end{align}
By performing the see-saw mechanism, the mass matrix for the active neutrino is given by
\begin{align*}
 m_{light}&\approx -m_{D}^{T}M_{M}^{-1}m_{D}\\
 &\approx     
\frac{v_{2}^{2}}{\sqrt{2}h_{1}v_{S}}\left(
\begin{array}{ccc}
(h_{2e}^{\nu e})^{2}+(h_{2\mu}^{\nu e})^{2}\rho&*&*\\
h_{2e}^{\nu e}h_{2e}^{\nu\mu}+h_{2\mu}^{\nu e}h_{2\mu}^{\nu\mu}\rho&
(h_{2e}^{\nu\mu})^{2}+(h_{2\mu}^{\nu\mu})^{2}\rho&
*\\
h_{2e}^{\nu e}h_{2e}^{\nu\tau}+h_{2\mu}^{\nu e}h_{2\mu}^{\nu\tau}\rho&
h_{2e}^{\nu\mu}+h_{2e}^{\nu\tau}\rho&
(h_{2e}^{\nu\tau})^{2}+(h_{2\mu}^{\nu\tau})^{2}+(h_{2\mu}^{\nu\tau})^{2}\rho
\\
\end{array} 
\right) ,
\end{align*}
where 
\begin{equation}
 \frac{v_{2}^{2}}{\sqrt{2}h_{1}v_{S}}\approx\frac{m_{\tau}^{2}}{M_{1}}.
\end{equation}
Assuming that $v_{2}\approx m_{\tau}$ and $v_{S}\approx 10^{10}\mathrm{GeV}$ yields that the light neutrinos have masses at the order of eV, and the squared mass differences will be fixed by the $h_{2e(\mu)}^{\nu i}$ Yukawa couplings. Since $\mathrm{Rank}(m_{light})=0$, the lightest of the three active neutrinos is massless, which would be $v_{L}^{1}$ for normal ordering or $v_{L}^{3}$ for inverted ordering\cite{NuFIT}. 

The $\rho$-parameter is defined as
\begin{equation}
    \rho=\frac{M_{1}}{M_{2}}=\frac{h_{1}}{h_{2}},
\end{equation}
where we assume the hierarchy $M_{1}<M_{2}<M_{3}$ (although the active neutrino masses do not depend on $M_{3}$). Thus, in order to make the model consistent to the neutrino oscillation data \cite{NuFIT}, we take
\begin{equation}
    \rho=0.5,\;\;\; |h_{2}|\approx 0.1,
\end{equation}
so that the right-handed neutrinos obey the mass hierarchy
\begin{equation}
    M_{1}\approx 4\times 10^{10}\mathrm{GeV}= 0.5 M_{2}.
\end{equation}

Considering complex Yukawa parameters, the model provides a mechanism to study the Baryon Asymmetry of the Universe (BAU) that can be obtained through the leptonic asymmetry generated from the right-handed neutrinos. This remaining right-handed neutrinos are coupled to sphalerons in the electroweak phase, thus producing baryogenesis. The Davidson-Ibarra bound \cite{davidsonibarra}
\begin{equation}
    k_{1}=\frac{\Gamma_{D1}}{H|_{T=M_{1}}}=\frac{v_{2}^{2}(h_{2e}^{\dagger}h_{2e})_{11}}{m_{*}M_{1}},
\end{equation}
satisfies the condition $k_{1}<1$ to maintain the leptonic asymmetry and to avoid washout. Assuming that $m_{*}\approx 1\times 10^{-3}\mathrm{eV}$, $v_{2}\approx m_{\tau}$ and $|h_{2e}|\approx 0.1$, then the mass for the lightest right-handed neutrino is
\begin{equation}
    M_{1}>m_{\tau}(h_{2e}^{\dagger}h_{2e})_{11}\times 10^{3}\times 10^{9}\mathrm{GeV}
    \approx 1\times 10^{10}\mathrm{GeV},
\end{equation}
which is a value suited to explain both leptogenesis and see-saw mechanism type I.


\section{Axion phenomenological aspects}

For linear and non-linear realizations of the QCD axion models, the most general bosonic Lagrangian, including only the NLO corrections, is given by \cite{kaplan}
\begin{equation}
{\cal L}_{eff} = {\cal L}^{LO} + \delta {\cal L}^{bosonic}_a,
\end{equation}
where now the leading order Lagrangian ${\cal L}^{LO}$ is the model one plus the axion kinetic term,
\begin{equation}
{\cal L}^{LO} = {\cal L}_{model} +\frac{1}{2}\partial^{\mu} a \partial_\mu a,
\end{equation}
and the NLO effective Lagrangian is given by five-dimension operators that can be written as
\begin{equation}
\Delta{\cal L}^{bosonic}_a =  -\frac{1}{4}g_{agg} a G\tilde{G} - \frac{1}{4} g_{a\gamma\gamma} a F\tilde{F}  ,
\end{equation}
where $F, G$ are the field strength tensors and
the axion-gauge couplings are given in terms of the anomaly coefficients
which have been computed in \cite{nardi,nardi2,anomaloNE} for different representations in the standard invisible axion models
\begin{equation}
g_{a\gamma\gamma} =\frac{m_a}{eV}\frac{2.0}{10^{10} \mathrm{GeV}} \left(\frac{E}{N} - 1.92(4)\right).
\label{gaff}
\end{equation}
In these equations $E$, $N$ denote the electromagnetic and color anomaly coefficients, respectively, for a given fermion content $f$
\begin{align}
N &= \sum_{f}\left(X_{f_L} - X_{f_R} \right) T(C_{f}^{SU(3)}), \nonumber \\
E &= \sum_{f}\left(X_{f_L} - X_{f_R} \right) {\cal Q}^2_f ,
\end{align}
where $X_{f_L}$ and $X_{f_R}$ are the PQ charges of the left- and right-handed  components of a given fermion representation 
$f$, ${\cal Q}_f$ is the $\mathrm{U}(1)_{Q}$ electromagnetic charge, and the $T(f)$ factor is the Dynkin index in the given representation which is given by $T(C_f)\delta_{ab}= Tr (T^a_f T^b_f)$ where $a, b$ are the group indices.  To calculate the decay of the axion into two photons it is necessary to estimate the effective coupling constant $g_{a\gamma\gamma}$ which depends on the $N/E$ ratio between anomaly coefficients.
It has been  computed for different representations of exotic fermions in the
standard invisible axion models in \cite{nardi,nardi2}.  
The numerical value in (\ref{gaff}) differs from the usual one 
 $E/N-1.92$ \cite{anomaloNE}, due to  higher order chiral corrections and in \cite{belen} is obtained
\begin{equation} 
g_{a\gamma\gamma} =\frac{m_a}{eV}\frac{2.0}{10^{10} \mathrm{GeV}} \left(\frac{E}{N} - 2.03\right).
\end{equation}
The ratio $E/N$ for this model takes the form
\begin{equation}
\frac{E}{N}=\frac{1}{3}\frac{6x_2-2x_\chi+3x_S}{-3x_1+4x_2-x_\chi} \approx \frac{3}{4},
\end{equation}
where we have taken the approximations for $x_1$, $x_2$, $x_\chi$ in (\ref{xapprox}) and $\sin\beta = 1$. 

The called cosmological Domain Wall (DW) problem presented in this class of models \cite{domainwall1} is associated to the fact that the energy density of the DW will largely overshoot the critical density of the Universe. The DW problem is avoided if ${\cal N}_{DW} = 1$ \cite{domainwall2}, then we might
want to consider this specific value as an additional feature for axion models. The ${\cal N}_{DW}$ is defined through the expression ${\cal N}_{DW}=N/x_S$. So,  the value that we find for the color anomaly coefficient  is 
$N=4 x_S$, then ${\cal N}_{DW} = 4$  and the model can not avoid the formation of DW.

In this model, active neutrinos acquired masses through the see saw mechanism type I and we had to choose the mass of the Majorana neutrinos of the order of $10^{10} \mathrm{GeV}$ which implies that $v_S$ is of the order of $10^{10} \mathrm{GeV}$. On the other hand, the axion decay constant is related to the Peccei Quinn symmetry breaking scale through the relation $f_a{\cal N}_{NW} = v_S$. Implies that $f_a \approx 10^{10} \mathrm{GeV}$ for this model and the axion mass and the effective axion-photon coupling  are
\begin{align}
m_a&=6\times 10^{-4} eV, \nonumber \\
|g_{a\gamma\gamma}| &=1.72\times 10^{-13}\times\left(2.03-0.75\right)\mathrm{GeV}^{-1}\nonumber \\
&\approx 2.20\times 10^{-13} \mathrm{GeV}^{-1}.
\end{align}

The absence of a significant signal above background in the data provides a leading limit on the axion-photon coupling strength
 $\mid g_{a\gamma\gamma}\mid <6.6\times 10^{-11}\mathrm{GeV}^{-1}$ 
 ($95\%$ C.L.)
, similar to the most restrictive astrophysical bounds \cite{cast}. 
ALPS II \cite{ALP} plans to explore
that region for a broader range of masses. The ADMX experiment had explored the  region of axion-photon favoured 
by the invisible axion \cite{ADMX}. Axion photonic couplings  are also being searched by the International Axion Observatory IAXO \cite{IAXO}.
The expected sensitivities for ALPS-II \cite{ALP}, IAXO \cite{IAXO}, ADMX \cite{ADMX2} and MADMAX \cite{MADMAX} are depicted to improve the bounds for $g_{a\gamma\gamma}$.

In the non-linear formalism, the interaction of the axion with the fermionic sector appears from the Yukawa Lagrangian with  bigger dimension scalar operators  and, after symmetry breaking, the NLO five dimension operators can be written as
\begin{align}
\Delta{\cal L}&=\frac{\partial_\mu a}{f_a} \sum_{f=u,d,e}\bar{f}^i\gamma_\mu 
(V_L^{(f)\dagger}X_{f_L}V_L^{(f)}P_L 
+ V_R^{(f)\dagger}X_{f_R}V_R^{(f)} P_R)_{ij} f^j \nonumber \\
&=\frac{\partial_\mu a}{2 f_a} \sum_{f=u,d,e}\bar{f}^i\gamma_\mu  (V_{ij L}^f P_L+ 
V_{ij R}^f P_R) f^j,
\end{align}
where $V^f$ and $A^f$ for the $f$ fermion are
\begin{equation}
V^f_L=V_L^{(f)\dagger}X_{f_L}V_L^{(f)}, \quad V^f_R =V_R^{(f)\dagger}X_{f_R}V_R^{(f)},
\end{equation}
with the associated matrices for the PQ charges  for up, down quarks and charged leptons, respectively, giving by
\begin{align}
X_{u_L}&=X_{d_L}=(0,1,0),\nonumber \\
X_{u_R}&=(1,0,1), \nonumber \\
X_{d_R}&=(1,1,-1),\nonumber \\
X_{e_L}&=\left(-\frac{3}{2}, -\frac{3}{2},\frac{1}{2}\right), \nonumber \\
X_{e_R}&=\left(-\frac{1}{2},\frac{5}{2}, -\frac{1}{2}\right),
\end{align}
when $x_1= -c_\beta^2$, $x_2=s_\beta^2 $, $x_\chi\approx 0 $ and $x_S=1 $. 

In order to study flavour changing neutral currents $sd$ it is necessary to calculate the expression for the matrices $V_{L}^d$ and $V_{R}^d$ which can written as $3\times 3$ matrix in the following form
\begin{equation}
V^d_L\approx \left(
\begin{array}{ccc}
(s_{\theta_L}^{(d)})^2 & - s_{\theta_L}^{(d)} & 0 \\
- s_{\theta_L}^{(d)} & 1 & 0 \\
0 & 0 & 0
\end{array}
\right), \;\;\;
V^d_R\approx \left(
\begin{array}{ccc}
1 & 0 & 0 \\
 0 &  1 & 0  \\
0 & 0 & - 1
\end{array}
\right).
\end{equation}
For the flavour changing neutral currents in the charged lepton sector we have 
\begin{equation}
\label{eq:axion-lepton-couplings}
V^e_L \approx \frac{1}{2}\left(
 \begin{array}{ccc}
 -3 & 0  & \frac{3}{\sqrt{2}} \frac{m_\mu}{m_\tau} \\
0 & -3 &  -4 \frac{m_\mu}{m_\tau} \\
\frac{3}{\sqrt{2}} \frac{m_\mu}{m_\tau} & -4\frac{m_\mu}{m_\tau}  & 1
\end{array}
\right), \quad
V^e_R \approx \frac{1}{2}\left(
\begin{array}{ccc}
 -1 & -5 \frac{m_e}{m_\mu} & 0    \\
 -5 \frac{m_e}{m_\mu}  & 5 & -\frac{1}{\sqrt{2}}\frac{m_\mu}{m_\tau}  \\   0 & -\frac{1}{\sqrt{2}}\frac{m_\mu}{m_\tau} & -1
\end{array}
\right).
\end{equation}

The axion effective couplings with flavour changing predicted by the model are giving by 
\begin{equation}
\Delta{\cal L}\approx- i \frac{m_s}{f_a}s_{\theta_L}^{(d)} a  
\bar{d}P_R s - 
i \frac{m_\mu}{f_a} \frac{5 m_e}{2m_\mu}  a \bar{e} P_L \mu  
+ i \frac{m_\tau}{f_a}\frac{3 m_\mu}{2\sqrt{2}m_\tau}  a\bar{e}P_R \tau -
 i \frac{m_\tau}{f_a}\frac{2m_\mu}{m_\tau} a \bar{\mu}P_R\tau.
\end{equation}
The matrices $V^e_L$ and $V^e_R$ in (\ref{eq:axion-lepton-couplings}) show the contributions of the two chiralities to the flavour changes mediated by the axion. The flavour changing $\mu-\tau$ receives contributions from the left- and right-handed chiralities, but only the right-handed one is kept; regarding $e-\tau$ flavour changing only the right-handed chirality contributes, while for $e-\mu$ the left-handed does so. 

The decay width  for flavour changing through the axion $K\to\pi$ is giving by
\begin{equation}
\Gamma(K^+\to \pi^+ a) = \frac{m_k^3}{64 \pi} \left(\frac{s_{\theta_L}^{(d)}}{f_a}\right)^2 \!\!\!
\mid f_0(q^2)\mid^2 \left(1- \frac{m_\pi^2}{m_K^2}\right)^3, \!\!\!
\end{equation}
where $m_{K}, m_\pi$ are the kaon and pion masses, and $f_0$ is the scalar form factor defined by the matrix element $\left<\pi^+\mid\bar{d}s\mid K^+\right>$ $=(m_K^2-m_\pi^2)f_0(q^2)$ which we take the order of 1 \cite{f0}. Using the $ BR(K^+\to \pi^+ \nu\bar{\nu})\approx \left(17.3^{+1.15}_{-1.05}\right) \times 10^{-11}$ \cite{kpibound} bound from the  E949 \cite{NA62} and E787 \cite{NA63} experiments we  can reinterpret it to find a bound to $BR(K^+\to \pi^+ a)$. Taking into account that the SM prediction is $BR(K^+\to \pi^+ \nu\bar{\nu})=(9.11\pm 0.72)\times 10^{-11}$ \cite{kpiSM}, we use the the upper bound
$BR(K^+\to \pi^+ a ) < 7.3 \times 10^{-11}$ \cite{kpa}, which gives 
\begin{equation}
s_{\theta_L}^{(d)} < 2.7 \times 10^{-2},
\label{eq:sdL}
\end{equation}
for $f_a=10^{10} \mathrm{GeV}$. 

From the non-linear effective lepton Lagrangian the decay widths for flavour changing are \cite{MADMAX}
\begin{align}
\Gamma(\mu\to e a) &=\frac{m_\mu}{16\pi}    \left( \frac{5m_e}{f_a}\right)^2, \nonumber\\
\Gamma(\tau \to \mu a)& =\frac{m_\tau}{16\pi}\left(\frac{4 m_\mu}{f_a}\right)^2,  \\
\Gamma(\tau \to e a) &=\frac{m_\tau}{16\pi}\left(\frac{3m_\mu}{\sqrt{2}f_a}\right)^2.	\nonumber 
\end{align}
By using  $\tau_{\tau}=(290.3\pm 0.5)\times 10^{-15}$ s and $\tau_{\mu}=2.19\times 10^{-6}$ \cite{PDG} and taking  $f_a=10^{10} \mathrm{GeV}$,  we find the branching ratios for the flavor changing of the lepton sector which are the order of
\begin{align}
BR(\mu\to e a) &\approx  4.5\times 10^{-10}, \nonumber\\
BR(\tau \to \mu a) &\approx 2.8 \times 10^{-11}, \\
BR(\tau \to e a) &\approx  7.7 \times 10^{-12}.\nonumber
\end{align}

\section{Conclusions}

The SM leaves several issues unexplained, {\it e.g.}, the fermion mass hierarchy, the strong $CP$-problem, the masses of neutrinos, baryogenesis, among others. A $\mathrm{U}(1)_{X}$ gauge extension, anomaly free, which distinguishes among families is introduced. The PQ symmetry is used to generate the accurate Ansatz for the mass matrices in order to understand the fermion mass hierarchy without using fine tuning in the Yukawa sector. In order to make the axion invisible, it is decoupled at low energies from the $Z_\mu$ and $Z'_\mu$ would-be Goldstone bosons, $G_{Z}$ and $G_{Z'}$, respectively. Then, the axion only couples to the right-handed neutrinos. The  PQ-symmetry found allows to solve the strong $CP$-problem where the QCD anomaly is different from zero. 

The model requires two Higgs doublets; one of them gives masses at tree-level to the top and charm quarks, where the latter gets its mass through a see-saw mechanism with the new $T$ exotic quark so that its mass becomes small respect to the top quark, due to the mutual cancellation between Yukawa couplings, as is observed in \eqref{eq:upsectormass}. The second Higgs doublet gives masses to the bottom quark and the $\mu$ and $\tau$ charged  leptons at tree-level. The up and down quarks and the electron are massless up to radiative corrections at 1-loop. Finite masses are generated through the introduction of a scalar singlet $\sigma$ which propagates inside the loop and, in the particular choice $v_1\gg v_2$  cause the complete decoupling of the bottom quark from the other two down quarks due to $\theta\sim\beta\sim\pi/2$, which seems to be a reasonable choice in order to the charged leptons couple to the Higgs in the same way as the SM predicts. In fact, it is possible to generate finite masses to the lightest fermions through the effective operators invariant under $U(1)_{X}\times Z_{2}$ which contribute to the correct Ansatz for the fermion mass matrices \cite{alvarado2021}.


The masses of the active neutrinos are given throughout the see-saw  mechanism type I, where the $\nu_{R}^{e,\mu,\tau}$ neutrinos  acquire masses due to VEV of the $S$ singlet scalar field  at the PQ scale, at the order of $\left<S\right>\sim 10^{10} \mathrm{GeV}$. On the other hand, the scale for the masses of the active neutrinos is set by the following relation, 
\begin{equation}
 \frac{v_2^2}{\sqrt{2}h_{1}v_{S}}\approx \frac{m_\tau^2}{M_1}.
\end{equation}
The $\nu_{R}^{1}$ right-handed neutrino with mass $M_{1}\sim 10^{10} \mathrm{GeV}$ can be also a candidate to take part of the $CP$-baryonic violation through leptogenesis. In this case, it is possible to have $\Gamma_{D1}<H\mid_{T=M_1}$ to avoid the washout. 

The Yukawa Lagrangians in \eqref{eq:Lagrangian-quark-sector} and \eqref{eq:Leptonic-Lagrangian} of the present model were generated by $Z_2$ symmetry in Ref. \cite{Mantilla}, which together to the Higgs potential in  \eqref{eq:scalar-potential} in turn yield the constrains as a set of coupled equations shown in the \ref{app:PQ-charges} to get the PQ charges for the fermions. However, these equations have modified the Yukawa Lagrangian for the down-quark sector by breaking the universality among the down-quarks of the model that existed in the Ref. \cite{Mantilla}. In particular, this modification isolates the bottom quark from being mixed with the lighter down quarks at tree level, which in turn are mixed by the angle $\theta^{(d)}_{L}$ through radiative corrections, constrained to be $s_{\theta_L}^{(d)}<0.027$ by the experiments E787 and E949 on the process $BR(K^+\to \pi^+ a)$.

The model introduced in the present article includes an extension to the gauge group of the SM with a new abelian non-universal $\mathrm{U}(1)_{X}$, as well as a global abelian group $\mathrm{U}(1)_{PQ}$ to ease the appropriate zeros in the mass matrices in order to get a good mass hierarchy, and also to address the strong-CP problem with the new scalar $S$. Both interactions produce new effects, specially important on the quark sector which suggest three criteria to be taken into account in order to get new physics consistent with the current experimental evidence: 1) the flavor changes in the down quark sector, precisely the rare decay $K\rightarrow \pi a$ due to the coupling to the axion $a$, 2) the Cabibbo angle of the CKM matrix, and 3) the non-universality of the neutral $Z'$ current. 
In addition, the requirement that the  abundance of axions is lower than the observed Cold Dark Matter implies that the axion decay constant is $f_a<(8.7 - 6.1)\times 10^{10} \mathrm{GeV}$ which is of the same order of the right-handed neutrino masses in order to get appropriate active neutrino masses in the present model. 

Considering the non linear effective Lagrangian for the leptonic sector with the axion, we predicted the following branching ratios 
\begin{align}
BR(\mu\to e a) &\approx  4.5\times 10^{-10}, \nonumber\\
BR(\tau \to \mu a) &\approx 2.8 \times 10^{-11}, \\
BR(\tau \to e a) &\approx  7.7 \times 10^{-12}.\nonumber
\end{align}
Finally, the effective coupling constant between axion and two photons is lower than $2.2\times 10^{-13}\mathrm{GeV}^{-1}$, in good agreement to the experiments, although the model does not solve the domain wall problem because the value obtained for ${\cal N}_{DW}$ is 4.

\appendix
\section{PQ charges}
\label{app:PQ-charges}


Under the $\mathrm{U}(1)_{PQ}$ symmetry, the scalar fields transform as:
\begin{align}
    \phi_{1}&\rightarrow e^{ix_{1}\alpha}\phi_{1},\qquad
    \phi_{2}\rightarrow e^{ix_{2}\alpha}\phi_{2},\nonumber\\
    \chi&\rightarrow e^{ix_{\chi}\alpha}\chi,\qquad
    S\rightarrow e^{ix_{S}\alpha}S.
\end{align}
and the current associated to the PQ transformation is given by
\begin{equation}
    J_{\mu}^{PQ}=x_{S}v_{S}i\partial_{\mu}\zeta_{S}+x_{\chi}v_{\chi}i\partial_{\mu}\zeta_{\chi}
    +x_{2}v_{2}\partial_{\mu}\eta_{2}+x_{1}v_{1}\partial_{\mu}\eta_{1},
    \label{eq:PQ-current}
\end{equation}
which must be orthogonal to the neutral currents at low energies
\begin{align}
\fl
 \left<J_{\mu}^{PQ}|G_{Z}\right>=x_{1}v_{1}s_{\beta}\left<\partial_{\mu}\eta_{1}|\eta_{1}\right>+x_{2}v_{2}c_{\beta}\left<\partial_{\mu}\eta_{2}|\eta_{2}\right>
 +\frac{m_{Z'_{\mu}}}{m_{Z_{\mu}}}s_{Z}x_{\chi}v_{\chi}\left<\partial_{\mu}\zeta_{\chi}|\zeta_{\chi}\right>=0,\\
 \fl
 \left<J_{\mu}^{PQ}|G_{Z'}\right>=-\frac{2v_{1}^{2}}{v_{\chi}}x_{1}\left<\partial_{\mu}\eta_{1}|\eta_{1}\right>-\frac{v_{2}^{2}}{v_{\chi}}x_{2}\left<\partial_{\mu}\eta_{2}|\eta_{2}\right>
 +x_{\chi}v_{\chi}\left<\partial_{\mu}\zeta_{\chi}|\zeta_{\chi}\right>=0.
\end{align}
The last expression can be simplified using $s_{Z}$ from \cite{Mantilla}
\begin{equation}
 \frac{m_{Z'}}{m_{Z}}s_{Z}\approx\left(\frac{2v_{1}^{2}+v_{2}^{2}}{v^{2}}\right)\frac{v}{v_{\chi}}.
\end{equation}
Therefore, the following equations describe the restrictions on the PQ charges for the scalar sector
\begin{align}
   0&=v_{1}^{2}x_{1}+v_{2}^{2}x_{2}+(2v_{1}^{2}+v_{2}^{2})x_{\chi},\nonumber\\
   0&=2v_{1}^{2}x_{1}+v_{2}^{2}x_{2}-v_{\chi}^{2}x_{\chi}.
    \label{eq:AxionNorm}
\end{align}
In addition, the $\lambda_{14}$ term in the scalar potential (\ref{eq:scalar-potential}) generates the following equation
\begin{equation}
x_{S}=-x_{1}+x_{2}+x_{\chi}.
\label{eq:AxionHiggs}
\end{equation}
By using \eqref{eq:AxionNorm} and \eqref{eq:AxionHiggs}), it is possible to write the PQ charges of the scalar fields as
\begin{align}
    x_{1}&=\frac{(-2v^{2}+v_{2}^{2}-v_{\chi}^{2})v_2^2}{4v^{4}-5v^2v_2^2+2v_{2}^{4}+v^{2}v_{\chi}^{2}}x_S,\nonumber \\  
    x_{2}&=\frac{4v^{4}-6v^2v_{2}^{2}+2v_2^4+v_1^2v_{\chi}^{2}}{4v^{4}-5v^2v_2^2+2v_{2}^{4}+v^{2}v_{\chi}^{2}}x_S,\nonumber\\    
    x_{\chi}&=\frac{-v_{1}^{2}v_{2}^{2}}{4v^{4}-5v^2v_2^2+2v_{2}^{4}+v^{2}v_{\chi}^{2}}x_S,
\end{align}
and, taking into account the hierarchy between VEVs $v_{\chi}\gg v_1, v_2$, the above expressions can be approximated to 
\begin{align}
    x_{1}&\approx -c^2_\beta x_S,\nonumber \\  
    x_{2}&\approx s^2_\beta x_S,\nonumber\\    
    x_{\chi}&\approx -s^2_\beta c^2_\beta \frac{v^2}{v_\chi^2}x_S.
    \label{xapprox}
\end{align}
In the limit where $v_S\gg v_\chi, v_1, v_2$,
\begin{equation}
x_1 s_\beta^2+x_2c_\beta^2 =0,
\end{equation}
which is a similar result for models with TDHM and PQ charges \cite{DFS1981}.

These charges with the Ansatze for the fermion mass matrices coming from the Yukawa Lagrangians of (\ref{eq:Lagrangian-downquarkS-sector}) and (\ref{eq:Leptonic-Lagrangian}) define the set of constrains on the PQ charges of the fermions. Now, the idea is to express all charges of the quark and lepton sectors based on these four values. With the Yukawa Lagrangian (\ref{eq:Lagrangian-downquarkS-sector}) and the scalar potential (\ref{eq:scalar-potential}), it is possible to write down the following restrictions on the PQ charges for the up quark sector

\begin{subequations}\label{eq:PQ_upquark}
\begin{align}
-x_{q_L^1}-x_2+x_{u_R^2}&=0,   &-x_{q_L^1}-x_2+x_{T_R}&=0,     
\tag{\theequation a-b} \\
-x_{q_L^2}-x_1+x_{u_R^2}&=0,  &-x_{q_L^2}-x_1+x_{T_R}&=0,                     \tag{\theequation c-d} \\
-x_{q_L^3}-x_1+x_{u_R^1}&=0 , &-x_{q_L^3}-x_1+x_{u_R^3}&=0,
\tag{\theequation e-f} \\
-x_{T_L}+x_\chi+x_{u_R^2}&=0, &-x_{T_L}+x_\chi+x_{T_R}&=0,
\tag{\theequation g-h}\\
-x_{T_L}+x_\sigma+x_{u_R^1}&=0.
\tag{\theequation i}
\end{align}
\end{subequations}

From (\ref{eq:PQ_upquark}a) to (\ref{eq:PQ_upquark}d) it is possible to infer that $x_{u_R^2}=x_{T_R}$, while (\ref{eq:PQ_upquark}e) and
(\ref{eq:PQ_upquark}f) imply that $x_{u_R^1}=x_{u_R^3}$. Leaving $x_{q_L^1}$ and $x_{q_L^3}$ as free charges, and also by adding the last restrictions, the solutions of the set of equations can be expressed as
\begin{subequations}\label{eq:PQ_downquark}
\begin{align}
    x_{q_L^{2}}&=-x_{1}+x_{2}+x_{q_L^{1}},
    &x_{u_R^{1}}&=x_{q_L^{1}}+x_{1},\\
    x_{u_R^{2}}&=x_{u_R^3}=x_{q_L^{3}}+x_{2},
    &x_{T_R}&=x_{q_L^{1}}+x_{2},\\
    x_{T_L}&=x_{\chi}+x_{q_L^{1}}+x_{2},
    &x_{\sigma}&=x_{\chi}+x_{2}-x_{1}=x_{S}.
    \label{eq:xsigma}
\end{align}
\end{subequations}
The values of the PQ-charges in (\ref{eq:PQ_upquark}i)  allow the $T_{L}\sigma U_{R}^{1}$-vertex which is used to induce radiative corrections at $1$-loop level and generate the up quark mass.

In the same way, the Lagrangian in (\ref{eq:Lagrangian-quark-sector}) enforces the following restrictions for the down-sector
\begin{subequations}\label{eq:PQ_updown}
\begin{align}
-x_{q_L^1}+x_1+x_{J_R^a}&=0,    &-x_{q_L^2}+x_2+x_{J_R^a}&=0,    
 \tag{\theequation a-b} \\
-x_{q_L^3}+x_2+x_{D_R^3}&=0,    &-x_{J_L^a}-x_{\chi}+x_{J_R^{a}}&=0,                     \tag{\theequation c-d} \\
-x_{J_L^a}-x_{\sigma}+x_{D_R^1}&=0, & -x_{J_L^a}-x_{\sigma}+x_{D_R^2}&=0.
\tag{\theequation e-f}
\end{align}
\end{subequations}
where $a,b=1,2$. The $J_{L}^n\sigma D_{R}^a $ couplings in  (\ref{eq:PQ_updown}e-f) are necessary to give masses to the down and strange quarks at $1$-loop level. From (\ref{eq:PQ_updown}a-f), it is obtained that
\begin{align}
    x_{J_R^{a}}&=x_{q_L^{1}}-x_{1},\label{eq:pq1}\\
    x_{D_{R}^{3}}&=x_{2}-x_{q_{L}^{3}},\\
    x_{J_L^{a}}&=x_{J_R^{a}}-x_{\chi}=x_{q_{L}^{1}}-x_{1}-x_{\chi},\label{eq:pq2}\\
      x_{D_{R}^{1}}&=x_{D_{R}^{2}}=-2x_{1}+x_{2}+2x_{q_L^{1}}-x_{q_{L}^{3}}\label{eq:pq3},
  \end{align}
where the value of the $x_{\sigma}$ charge given in (\ref{eq:xsigma}) has been employed.
 

\begin{table*}
\centering
\begin{tabular}{ccc|ccccc|}
\hline\hline
$PQ$-label	&$PQ$-charge&&	$PQ$-label&$PQ$-charge	\\ \hline 
\multicolumn{5}{c}{SM Fermionic Isospin Doublets}	\\ \hline\hline
$x_{q_L^1}$	&	&&
$x_{\ell_L^{e}}$	&$-\dfrac{x_{S}}{2}-x_{2}$	\\
$x_{q_L^2}$	&$-x_1+x_2+x_{q_L^1}$	&&
$x_{\ell^{\mu}_{L}}$	&$-\dfrac{x_{S}}{2}-x_{2}$		\\
$x_{q_L^3}$	&	&&
$x_{\ell^{\tau}_{L}}$	&$-x_{1}+\dfrac{x_{S}}{2}+x_{\chi}$	\\   \hline\hline

\multicolumn{5}{c}{SM Fermionic Isospin Singlets}	\\ \hline\hline 
\begin{tabular}{c}$x_{U_{R}^{1,3}}$\\$x_{U_{R}^{2}}$\\$x_{D_{R}^{1,2}}$\\$x_{D_{R}^{3}}$\end{tabular}	&
\begin{tabular}{c}$x_{1}+x_{q_L^3}$\\$x_{2}+x_{q_L^1}$\\$x_2-2x_{1}+x_{q_L^1}$\\$-x_{2}+x_{q_L^1}$\end{tabular}	&&	
\begin{tabular}{c}$x_{e_{R}^{e,\tau}}$\\$x_{e_{R}^{\mu}}$\end{tabular}	&	
\begin{tabular}{c}$-x_{1}-x_{2}+x_{\chi}+\dfrac{x_{S}}{2}$\\$-\dfrac{x_{S}}{2}-2x_{2}$\end{tabular}\\   \hline \hline 

\multicolumn{2}{c}{Non-SM Quarks}	&&	\multicolumn{2}{c}{Non-SM Leptons}	\\ \hline \hline
\begin{tabular}{c}$x_{T_{L}}$\\$x_{T_{R}}$\end{tabular}	&
\begin{tabular}{c}$x_{\chi}+x_{2}+x_{q_L^1}$\\$x_{2}+x_{q_L^1}$\end{tabular}	&&
\begin{tabular}{c}$x_{\nu_{R}^{e,\mu,\tau}}$\\$x_{E_L}$\\$x_{E_R}$\end{tabular}	&	
\begin{tabular}{c}$-\frac{x_{S}}{2}$\\$-x_{1}-x_{2}+x_{\chi}-\dfrac{x_{S}}{2}$\\$-x_{1}-x_{2}-\dfrac{x_{S}}{2}$\end{tabular}\\
$x_{J^{1,2}_{L}}$ 	&$-x_{1}-x_{\chi}+x_{q_L^1}$	&&$x_{\mathcal{E}_{L}}$	&$-2x_{2}+\dfrac{x_{S}}{2}$	\\
$x_{J^{1,2}_{R}}$	&$-x_{1}+x_{q_L^1}$	&&$x_{\mathcal{E}_{R}}$	&$-2x_{2}+x_{\chi}+\dfrac{x_{S}}{2}$	\\ \hline \hline

\end{tabular}
\caption{Fermionic $PQ$-charge assignment according to the proposed Yukawa Lagrangian densities in \eqref{eq:Lagrangian-downquarkS-sector} and \eqref{eq:Leptonic-Lagrangian}.}
\label{tab:PQ-charges}
\end{table*}

By using the fact that the PQ current is axial, it is possible to set $x_{q_{L}^{1}}=x_{q_{L}^{3}}=0$ without loss of generality. 

For the charged leptonic sector, the restrictions followed from the Yukawa Lagrangian \eqref{eq:Leptonic-Lagrangian} are
\begin{subequations}
\begin{align}
    -x_{\ell_L^e}+x_2+x_{e_R^\mu}&=0, &-x_{\ell_L^\tau}+x_2+x_{e_R^e}&=0,\tag{\theequation a-b}\\
    -x_{\ell_L^e}+x_1+x_{E_R}&=0, & -x_{\ell_L^\mu}+x_2+x_{e_R^\mu}&=0,
    \tag{\theequation c-d}\\
   -x_{\ell_L^\tau}+x_2+x_{e_R^\tau}&=0, &-x_{\ell_L^\mu}+x_1+x_{E_R}&=0,
    \tag{\theequation e-f}\\
    -x_{E_L}-x_{\sigma}+x_{e_R^e}&=0, &-x_{\mathcal{E}_L}+x_{\sigma}+x_{e_R^\mu}&=0,\tag{\theequation g-h}\\
    -x_{E_L}+x_{\chi}+x_{E_R}&=0, &-x_{E_L}-x_{\sigma}+x_{e_R^\tau}&=0,
     \tag{\theequation i-j}\\
    -x_{\mathcal{E}_L}-x_{\chi}+x_{\mathcal{E}_R}&=0.    \tag{\theequation k}      
    \end{align}
    \label{eq:pq-leptonic}
\end{subequations}
From (\ref{eq:pq-leptonic}a-d) it is possible to see that $x_{\ell_L^{\mu}}=x_{\ell_L^{e}}$
and from (\ref{eq:pq-leptonic}b-d) it is obtained that $x_{e_R^{e}}=x_{e_R^{\tau}}$. 

Lastly, regarding the mass generation of neutrinos given by the Yukawa Lagrangian in \eqref{eq:Leptonic-Lagrangian}, the restrictions over the PQ-charges are
\begin{subequations}
\begin{align}
    -x_{\ell_L^e}-x_2+x_{\nu_R^{e}}&=0,
    &-x_{\ell_L^\mu}-x_2+x_{\nu_R^e}&=0,
    \tag{\theequation a-b}\\
     -x_{\ell_L^e}-x_2+x_{\nu_R^{\mu}}&=0,
    &-x_{\ell_L^\mu}-x_2+x_{\nu_R^{\mu}}&=0,
    \tag{\theequation c-d}\\ 
    -x_{\ell_L^e}-x_2+x_{\nu_R^{\tau}}&=0,
    &-x_{\ell_L^\mu}-x_2+x_{\nu_R^{\tau}}&=0,
    \tag{\theequation e-f}\\
    x_{\nu_{R}^{i}}+x_{S}+x_{\nu_{R}^{j}}&=0.\tag{\theequation g}
    \end{align}
    \label{eq:pq-leptonicn}
\end{subequations}
From (\ref{eq:pq-leptonicn} a, c ,d) it is seen that $x_{\nu_R^{e}}=x_{\nu_R^{\mu}}=x_{\nu_R^{\tau}}$. (\ref{eq:pq-leptonic} b, d, f) are equivalent because $x_{\ell_L^e}=x_{\ell_L^\mu}$. Therefore,
\begin{align}
 x_{\nu_R^{i}}&=x_{2}+x_{\ell_L^\mu}, \quad i=e,\mu,\tau,\\\label{eq:PQ_L}
 x_{\nu_{R}^{i}}&=-\frac{x_{S}}{2},\\ \label{eq:PQ_neutrino}
 x_{\ell_L^\mu}=x_{\ell_L^e}&=x_{\nu_R^{i}}-x_{2}=-\frac{x_{S}}{2}-x_{2}.
\end{align}

Finally, the whole set of the PQ charges that reproduce the same Yukawa Lagrangians given in \cite{Mantilla} are shown in the table (\ref{tab:PQ-charges}).

\section{Fermion mass matrices}
\label{app:Fermion-mass-matrices}
The matrices $M_{U}$, $M_{D}$ and $M_{E}$ in \eqref{eq:Up-mass-matrix}, \eqref{eq:Down-mass-matrix} and \eqref{eq:Electron-mass-matrix} are not Hermitian, so it is required to diagonalize its squares. The explicit form of the $4\times4$ matrix $M_{U}^{2}=M_{U}M_{U}^{\dagger}$ and the $5\times5$ matrix $M_{D}^{2}=M_{D}M_{D}^{\dagger}$ are shown below.

\begin{align} 
\label{eq:mu2}
M_{U}^{2}=\frac{1}{2} && \left(
\begin{array}{cc}
 v_{2}^{2}[(h_{2}^{T})_{1}^{2}+(h_{2}^{U})_{12}^{2}] &
*\\
v_{1}v_{2}[(h_{1}^{T})_{2}(h_{2}^{T})_{1}+(h_{1}^{U})_{22}(h_{1}^{U})_{12}] &v_{1}^{2}[(h_{1}^{T})_{2}^{2}+(h_{1}^{U})_{22}^{2}]\\
0 & 
0 \\
v_{2}v_{\chi}[(h_{2}^{T})_{1}(h_{\chi}^{T})+(h_{2}^{U})_{12}(h_{\chi}^{U})_{2}] & v_{1}v_{\chi}[(h_{1}^{T})_{2}(h_{\chi}^{T})+(h_{1}^{U})_{22}(h_{\chi}^{U})_{2}] \\
\end{array} \right. \nonumber 
\\ &&\vspace{1cm} \nonumber 
\\ \qquad \qquad 
&&\left.
\begin{array}{cccc}
*&
*\\
*&
*\\
v_{1}^{2}[(h_{1}^{U})_{31}^{2}+(h_{1}^{U})_{33}^{2}]&
*\\
0&
v_{\chi}^{2}[(h_{\chi}^{T})^{2}+(h_{\chi}^{U})_{2}^{2}]
\end{array}
\right). 
\end{align}

\begin{align}
\label{eq:md2}
M_{D}^{2}=\frac{1}{2} && \left(
\begin{array}{ccc}
v_{1}^{2}[(h_{1}^{J})_{11}^{2}+(h_{1}^{J})_{12}^{2}] &
*&
*\\
v_{1}v_{2}[(h_{1}^{J})_{11}(h_{2}^{J})_{21}+(h_{1}^{J})_{12}(h_{2}^{J})_{22}] &
v_{2}^{2}[(h_{2}^{J})_{21}^{2}+(h_{2}^{J})_{22}^{2}]
*\\
0 &
0&
v_{2}^{2}(h_{2}^{D})_{33}^{2}\\
v_{1}v_{\chi}[(h_{1}^{J})_{11}(h_{\chi}^{J})_{11}] &
v_{2}v_{\chi}[(h_{2}^{J})_{21}(h_{\chi}^{J})_{11}] &
0\\
v_{1}v_{\chi}[(h_{1}^{J})_{12}(h_{\chi}^{J})_{22}] &
v_{2}v_{\chi}[(h_{2}^{J})_{22}(h_{\chi}^{J})_{22}] &
0\\
\end{array} \right. \nonumber
\\ &&\vspace{1cm} \nonumber 
\\ \qquad \qquad 
&&\left.
\begin{array}{cccc}
* &&& 
* \\
* &&& 
* \\
* &&& 
*\\
v_{\chi}^{2}[(h_{\chi}^{J})_{11}^{2}] &&& 
*\\
0 &&&
v_{\chi}^{2}[(h_{\chi}^{J})_{22}^{2}]
\end{array}
\right). 
\end{align}

Finally, the $5\times5$ matrix $M_{E}^{2}=M_{E}M_{E}^{\dagger}$ is reduced to a $4\times4$ due to the exotic charged lepton $\mathcal{E}$ is disconnected from the rest of the system
\begin{small}
\begin{align} 
\label{eq:me2}
M_{E}^{2}=\frac{1}{2} && \left(
\begin{array}{ccc}
 v_{2}^{2}(g^{2e}_{e\mu})^{2}+v_{1}^{2}(g^{1}_{Ee})^{2} & 
v_{2}^{2}(g^{2e}_{e\mu}g^{2e}_{\mu\mu})+v_{1}^{2}(g^{1}_{E e }g^{1}_{E\mu}) &
0\\
v_{2}^{2}(g^{2e}_{e\mu}g^{2e}_{\mu\mu})+v_{1}^{2}(g^{1}_{Ee}g^{1}_{E\mu}) &	
v_{2}^{2}(g^{2e}_{\mu\mu})^{2}+v_{1}^{2}(g^{1}_{E\mu})^{2}  &
0\\
0	&	0	& v_{2}^{2}[(g^{2e}_{\tau e})^{2}+(g^{2e}_{\tau\tau})^{2}]\\
v_{1}v_{\chi}[g^{1}_{Ee}h^{\chi E}]	&	
v_{1}v_{\chi}[g^{1}_{Ee}h^{\chi E}] &
0\\
\end{array} \right. \nonumber 
\\ &&\vspace{1cm} \nonumber 
\\ \qquad \qquad 
&&\left.
\begin{array}{ccc}
v_{1}v_{\chi}[g^{1}_{Ee}h^{\chi E}]\\
v_{1}v_{\chi}[g^{1}_{Ee}h^{\chi E}]\\
0\\
v_{\chi}^{2}(h^{\chi E})^{2}
\end{array}
\right). 
\end{align}
\end{small}

\end{document}